\documentclass[journal,twoside,web]{IEEEtran}
\usepackage{cite}
\usepackage{amsmath,amssymb,amsfonts}
\usepackage{algorithmic}
\usepackage{graphicx}
\usepackage{algorithm,algorithmic}
\usepackage{hyperref}
\hypersetup{hidelinks=true}
\usepackage{textcomp}
\usepackage{subfigure}

\begin{document}
\title{The Invariant Zonotopic Set-Membership Filter for State Estimation on Groups}
\author{Tao Li, Yi Li, Lulin Zhang, and Jiuxiang Dong
\thanks{Tao Li, Yi Li, Lulin Zhang, and Jiuxiang Dong are with the College of Information Science and Engineering, the Liaoning Province Key Laboratory of Safe Operation Technique for Autonomous Unmanned Systems, and also with the State Key Laboratory of Synthetical Automation of Process Industries, Northeastern University, Shenyang 110819, China. (e-mail: litaoneu@stumail.neu.edu.cn; dongjiuxiang@ise.neu.edu.cn).}}

\maketitle

\newtheorem{assumption}{Assumption}[section]
\newtheorem{remark}{Remark}
\newtheorem{definition}{Definition}[section]
\newtheorem{property}{Property}[section]
\newtheorem{lemma}{Lemma}[section]
\newtheorem{theorem}{Theorem}[section]

\begin{abstract}
The invariant filtering theory based on the group theory has been successful in statistical filtering methods. However, there exists a class of state estimation problems with unknown statistical properties of noise disturbances, and it is worth discussing whether the invariant observer still has performance advantages. In this paper, considering the problem of state estimation with unknown but bounded noise disturbances, an Invariant Zonotopic Set-Membership Filter (InZSMF) method on groups is innovatively proposed, which extends the invariant filtering theory to the field of non-statistical filtering represented by set-membership filtering. Firstly, the InZSMF method transforms the state space from the traditional Euclidean vector space to the Lie group space to construct group affine discrete systems with unknown but bounded noise uncertainty defined by the zonotope on groups. Secondly, the nonlinear observer on the group is defined and the corresponding linearized estimation error is derived. Then, two observer gain tuning algorithms under the InZSMF method are proposed, respectively, the pole configuration method and the F-radius optimization method. Finally, through simulation experiments, it is shown that the InZSMF state estimation method is generally superior to the traditional Zonotopic Set-Membership Filter (ZSMF) state estimation method. Especially, when the initial estimations are imprecise, the convergence speed of state estimation, the accuracy of set-membership center estimation, and the average interval area of zonotopic estimation of the InZSMF method are significantly better than those of the ZSMF method. 
\end{abstract}


\section{Introduction}
\label{sec:introduction}

\IEEEPARstart{S}{tate} estimation is a method of estimating the internal state of a system by utilizing the prior model of the system and the available measurement data, which is an indispensable and important part of the automatic control system. However, due to the inherent random disturbances in the system and the limitations of the accuracy of the sensors acquiring the measurement data, uncertainties are inevitable, which is also an important reason affecting the accuracy of state estimation. Currently, there are two main ways to model uncertainty: a stochastic uncertainty that relies on the probability theory, and an unknown but bounded uncertainty that relies on the set-membership paradigm \cite{RAISSI20041771, COMBASTEL2015265, CONG2025111993}. The former usually assumes that the process and measurement noise of the system obeys some probability distribution, e.g., Gaussian noise, etc. The Kalman filtering method \cite{Kalman1960} is one of the commonly used methods to deal with the state estimation problem under Gaussian white noise. Of course, this requires that the statistical properties of the noise disturbance, such as mean and variance, need to be known in advance. However, sometimes the noise is unknown and information about the probability distribution it follows cannot be obtained. At this time, statistical filtering methods such as Kalman filtering are less suitable for use. Another state estimation method based on the set-membership paradigm does not rely on the exact statistical properties of the noise, but only utilizes the bounded information of the noise to deal with the state estimation problem. This method is also currently used in various fields \cite{9437974} such as localization \cite{10540277, 10685107, 8657799}, fault diagnosis \cite{XU2024111413,10670462,8624300, 10829798, 10678919,10184279}, event-triggered problems \cite{8985429, 10841464}, cyber-physical systems \cite{10124355}, etc. 

At present, both statistical filtering methods such as Kalman filtering and non-statistical filtering methods such as set-membership filtering are mostly based on the traditional Euclidean space. Interestingly, symmetry can be utilized to design asymptotic observers and invariant estimation errors for nonlinear systems, opening up new ideas for designing observers in non-Euclidean state spaces \cite{1184728}. The introduction of group theory led to the emergence of the invariant observer theory, which gradually developed into the invariant filtering theory \cite{7523335,4434662,barrau2018invariant,4700863}. A constructive method to find all the symmetry-preserving pre-observers is proposed and a theorem is proved that it is always possible to build an invariant observer whose linear tangent approximates is any linear asymptotic observer of the Luenberger type, around an equilibrium point \cite{4700863}. A $\theta$-invariant system as a generalization of symmetry-preserving systems is also proposed and used to construct a framework for observer design for homogeneous systems with arbitrary degrees \cite{10886075}. A large class of systems that can be transformed into group affine systems is discussed, and a construction method for projecting the group structures of a given system into an invariant filtering framework was proposed, systematically providing a method for constructing state space transformations from Euclidean spaces to the Lie groups \cite{9689974}. In the study of high-precision navigation problems, the Invariant Extended Kalman Filter (InEKF) method is proposed \cite{7523335}, which combines the invariant observer based on Lie group theory with the Extended Kalman Filter (EKF). By utilizing the geometric structure and dynamics of the state space, it has been shown that if the state space of a nonlinear system can be defined in the Lie group space, then the InEKF can obtain stronger estimation results than the traditional EKF \cite{7523335,4434662,barrau2018invariant}. This method is currently mainly applied in fields such as localization \cite{10697265, bourmaud2015continuous}, navigation\cite{9689974, 9812192, 9977251}, and simultaneous localization and mapping (SLAM) \cite{7523335,barrau2018invariant, 7812660}. And with its outstanding robustness, it has been truly applied in the aerospace field \cite{barrau2018invariant}. 

Invariant filtering theory has been increasingly used in the field of statistical filtering \cite{9444664, hartley2020contact, 9626926, 9812192, 8206066, 10697265}, however, the discussion in the field of non-statistical filtering is very limited and currently confined to interval estimation methods \cite{DAMERS2022110502, MERLINGE2024111688}. Due to the successful application of invariant filtering theory in the field of Kalman filtering, the extension of invariant filtering theory to the field of set-membership filtering has a high potential. In this paper, we will focus on the problem of state estimation for group affine systems with unknown but bounded noise disturbances, and construct a Zonotopic Set-Membership Filter (ZSMF) on the group space to study whether it has similar properties to the InEKF, which can improve the performance of the traditional set-membership estimation. To the best of our knowledge, this is also the first extension of invariant filtering theory to the field of Zonotopic Set-Membership. 

Fortunately, it is feasible to deal with the state estimation problem by using the ZSMF on the group, which is called the Invariant Zonotopic Set-Membership Filter (InZSMF) method in this paper. It is an exciting idea to extend the traditional ZSMF method from Euclidean vector space to Lie group space by combining the invariant filtering theory. From the results of simulation experiments, there is a significant improvement in the estimation accuracy, the area of the estimated set interval, and convergence speed compared with the traditional ZSMF method. The contributions of this paper are summarized as follows: 
\begin{itemize}
	\item[(1)] It is among the first attempts to develop the InZSMF method for dealing with the state estimation problem of group affine systems with unknown but bounded noise disturbances described by zonotope on the group. This method supplements the current invariant filtering theory and significantly improves the traditional set-membership filtering method.   
	\item[(2)] Within the InZSMF method, a very meaningful property is shown to remain, that is, the evolution of the invariant estimation error on the group approximately obeys the linear error dynamics in the corresponding vector space, which can help us to obtain a reachable set evolution form of the estimation error on the group. 
	\item[(3)] The observer gain tuning method is not unique in the InZSMF method, and two algorithms are proposed in this paper, namely, the pole configuration method and the F-radius optimization method. The former allows tuning the estimation performance, such as the overshooting, by designing the poles to meet engineering expectations. The latter reduces the error set in the estimation process by minimizing the F-radius metric with explicit geometrical interpretation to achieve more accurate online estimation. 
	\item[(4)] Simulation experiments show that the InZSMF state estimation method is generally superior to the traditional ZSMF state estimation method. Particularly, when the initial estimations are very imprecise, the state estimation convergence speed and the average interval area of zonotopic estimation of the InZSMF method are significantly higher than that of the ZSMF method, and the accuracy of state center estimation measured by the Root-Mean-Square Error (RMSE) can be improved by about $50\%$. 
\end{itemize}

The structure of this paper is organized as follows. Section \ref{sec:Preliminaries and Problem Formulation} introduces some preliminaries and presents the problems to be studied. Section \ref{sec:Invariant Zonotopic Set-Membership Filtering} describes the InZSMF method proposed in this paper and gives two algorithms for the design of the observer gain. In Section \ref{sec:simulation_experiment}, we conduct a simulation experiment study for a two-dimensional (2D) vehicle kinematic model to compare the performance of ZSMF and InZSMF methods. Finally, Section \ref{sec:Conclusion} summarizes this paper.

\section{Preliminaries and Problem Formulation}
\label{sec:Preliminaries and Problem Formulation}
\subsection{Preliminaries}
\label{subsec:preliminaries}
\begin{figure}[!t]
	\includegraphics[width=\columnwidth]{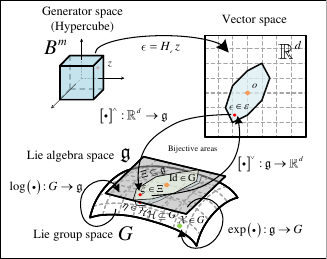}
	\caption{Illustration of the zonotope on the Lie group.}
	\label{fig1}
\end{figure}

This paper focuses on constructing new methods for state estimation by using group theory and set-membership filtering theory, and for ease of understanding, the notations and fundamentals that will be used in this paper will be introduced here. Let $ \mathbb{R}^{d} $, $ \mathbb{R}^{n \times m} $, and $ \mathbf{B}^{m} $ denote the set of $d$-dimensional vectors, the set of $n \times m$-dimensional matrices, and an $m$-dimensional hypercube, respectively. Let $ \mathcal{X} $, $ \eta $ , and $ \mathcal{H} $ denote the state variable, the error variable, and the set of the error described by zonotope on the Lie group, respectively. $ \xi $ and $ \varXi $ denote the error and the zonotopic set of the error on the Lie algebra, respectively. In the Euclidean vector space, $ \epsilon $ and $ \varepsilon $ denote the error variable and the zonotopic set of the error, respectively. $ z $ is the variable on the hypercube. $I_{n}$ denotes the identity matrix. The inverse, transpose and trace of matrix $A$ are denoted as $A^{-1}$, $A^{\top}$ and $tr(A)$ respectively. In the group theory, a set $ G $ is said to constitute a group over the binary operation $ \bullet $, denoted $ (G, \bullet) $, if a nonempty set $ G $ over the defined binary operation $ \bullet $ satisfies closure, the law of union, the existence of an identity element, and the existence of an inverse element. The above binary operation $ \bullet $ can also be called group multiplication. Moreover, if $ G $ is both an $n$-dimensional manifold and a group, and both its group multiplication mapping $ G \bullet G \to G $ and inverse element mapping $ G \to G $ are $ C^{\infty} $, then such a $ G $ is called an $n$-dimensional Lie group. A common Lie group is the matrix Lie group, i.e., a group structure consisting of rotation matrices under matrix multiplication. This is common in the description of kinematics such as the special orthogonal groups $SO2, SO3$ and the special Euclidean groups $SE2, SE3$. The tangent vectors can be made in different directions on an element $ \mathfrak{p} $ of the Lie group and the space formed by these tangent vectors is called a tangent space $ \mathcal{T}_{\mathfrak{p}} $. One of these tangent spaces at the identity element $ I_{d} $ is also called the Lie algebra space, denoted $ \mathfrak{g} $. A set of mapping relations between Lie groups and Lie algebras can be defined by using exponential mapping $ \exp(\cdot): \mathfrak{g} \to G $ and logarithmic mapping $ \log(\cdot): G \to \mathfrak{g} $, and it is easy to know that such a set of mappings satisfies bijection relation \cite{barrau2018invariant}. There exists a set of linear isomorphism mappings $ \left[ \cdot \right] ^ {\land}: \mathbb{R}^{d} \to \mathfrak{g} $ and $ \left[ \cdot \right] ^ {\lor}: \mathfrak{g} \to \mathbb{R}^{d} $ between Lie algebras and Euclidean vector spaces \cite{ 10697265, 9737005 } , and again such a set of mappings satisfies the bijection relation \cite{barrau2018invariant}. The mapping relations between the spaces are shown in Fig. \ref{fig1}. As for the characterization of uncertainty, this paper considers the case where the noise is unknown but bounded, and according to the set-membership filtering theory, the noise can be described as a set. For two sets $ X \in \mathbb{R}^{d} $ and $ Y \in \mathbb{R}^{d} $, the Minkowski sum is defined as $ X \bigoplus Y = \left\{ x + y \mid x \in X, y\in Y \right\} $. 

\begin{definition} \label{definition_zonotope}
	(from \cite{10685107}) An zonotope $\mathbf{Z} \subseteq \mathbb{R}^{d} $ of order $m$ can be expressed as $\mathbf{Z} = \left \langle p, H \right \rangle = \left\{ p + Hz, z \in \mathbf{B}^{m} \right\} $, where $p \in \mathbb{R}^{d}$ is the center of $\mathbf{Z}$, $H \in \mathbb{R}^{ d \times m }$ is the generator matrix of $\mathbf{Z}$, and $ \mathbf{B}^{m} $ is an $ m $-dimensional hypercube. 
\end{definition}

\begin{property} \label{property_zonotope_sum}
	(from \cite{10685107}) Given two zonotopes $\mathbf{Z}_{1} = \left \langle p_{1}, H_{1} \right \rangle $ and $\mathbf{Z}_{2} = \left \langle p_{2}, H_{2} \right \rangle $, their Minkowski sum is defined as $ \left \langle p_{1}, H_{1} \right \rangle \bigoplus \left \langle p_{2}, H_{2} \right \rangle = \left \langle p_{1}+ p_{2}, \begin{bmatrix} H_{1} ; H_{2} \end{bmatrix} \right \rangle $, where $p_{1}$ and $ p_{2} \in \mathbb{R}^{n}$ are the center elements of $\mathbf{Z}_{1} $ and $\mathbf{Z}_{2}$, respectively, $H_{1} \in \mathbb{R}^{d \times m_{1}}$ and $H_{2} \in \mathbb{R}^{ d \times m_{2} }$ are the generator matrices, and $ \begin{bmatrix} H_{1} ; H_{2} \end{bmatrix} $ denotes matrix concatenate. 
\end{property}

\begin{property} \label{property_zonotope_linear}
	(from \cite{10685107}) The linear mapping of a zonotope $\mathbf{Z}$ satisfies $ L\mathbf{Z} = \left \langle Lp, LH \right \rangle $, where $p \in \mathbb{R}^{d}$, $H \in \mathbb{R}^{ d \times m }$, and $L \in \mathbb{R}^{ s \times d }$. 
\end{property}

\begin{property} \label{property_zonotope_reduction}
	(from \cite{COMBASTEL2015265}) Given a zonotope $\mathbf{Z} = \left\{ p+Hz, z \in \mathbf{B}^{m} \right\} \subseteq \mathbb{R}^{d} $ and a fixed integer $ s \; (d< s <m) $, the matrix $H \in \mathbb{R}^{ d \times m }$ can be reduced to $\mathcal{R}_{s}(H) \in \mathbb{R}^{ d \times s }$ with the guarantee that $ \mathbf{Z} \in \left \langle p, \mathcal{R}_{s}(H) \right \rangle $. 
\end{property}

\begin{remark}
	The purpose of order reduction is to prevent the order of the generator matrix from increasing during the iterations of the set-membership filtering algorithm, which eventually exceeds the processing capacity of the computer. A common order reduction algorithm is as follows: $ \tilde{H} $ can be obtained by rearranging the columns of $H$ from largest to smallest in terms of the columns norm. Let $ \mathcal{R}_{s}(H) = \begin{bmatrix} H^{1} ; H^{2} \end{bmatrix}  $, where $ H^{1} $ is a matrix consisting of the first $ s-d $ columns of $ \tilde{H} $, and $ H^{2} $ is a diagonal matrix satisfying $ H_{i,i}^{2} = \begin{matrix} \sum_{j=s-d+1}^m \left| \tilde{H}_{i,j} \right| \end{matrix} $, $ i = 1, \cdots, d $. In the actual operation, when $ m > s $, the order reduction algorithm is used, otherwise the generator matrix is kept unchanged. 
\end{remark}

\begin{definition} \label{definition_F_raduis}
	(from \cite{COMBASTEL2015265}) For a zonotope $ \mathbf{Z} = \left \langle p, H \right \rangle \subseteq \mathbb{R}^{n} $, its F-radius is defined as the Frobenius norm of generator matrix $ H $: $ \Vert \left \langle p, H \right \rangle \Vert_{F} $ = $ \Vert H \Vert_{F} = \sqrt{tr(H^{\top}H)} $. 
\end{definition}

\begin{definition} \label{definition_covariance}
	(from \cite{COMBASTEL2015265}) The covariance matrix of a zonotope $ \mathbf{Z} = \left \langle p, H \right \rangle $ is defined as $ cov(\mathbf{Z}) = HH^{\top} $. 
\end{definition}

\begin{definition} \label{Group_affine}
	(from \cite{barrau2018invariant}) A dynamical system $ \mathcal{X}_{k+1} = f(\mathcal{X}_{k}, u_{k}) $ is called group affine system if $ f: G \to G $ satisfies the group affine property, which is that for all $ \mathcal{X}_{1}, \mathcal{X}_{2} \in G$ and $ \forall u $, it satisfies 
	\begin{equation} 
		f(\mathcal{X}_{1} \bullet \mathcal{X}_{2}, u) = f(\mathcal{X}_{1}, u) \bullet f(I_{d}, u)^{-1} \bullet f(\mathcal{X}_{2}, u) \label{group_affine_property}
	\end{equation}
	where $ u $ is the control input. 
\end{definition}

\begin{lemma} \label{Fundamental_property_of_invariant_filtering}
	(from \cite{barrau2018invariant}) The group affine property \eqref{group_affine_property} is equivalent to the fundamental property of invariant filtering that there exists a map $ g $ such that for $ \forall \mathcal{X}_{1}, \mathcal{X}_{2}, u$,  
	\begin{equation} 
		f(\mathcal{X}_{1}, u)^{-1} \bullet f(\mathcal{X}_{2}, u) = g(\mathcal{X}_{1}^{-1} \mathcal{X}_{2}, u). \label{fundamental_property_of_invariant_filtering} 
	\end{equation}
	Moreover, for each $ u \in \mathbb{R}^{n_{u}} $ there exists $ A \in \mathbb{R}^{d \times d} $ such that $ \forall \epsilon \in \mathbb{R}^{d} $, $ g(\exp([ \epsilon ]^{\land} ), u) = \exp([ A \epsilon ]^{\land}) + O(\left \| \epsilon \right \| ^ {2} ) \approx \exp([ A \epsilon ]^{\land}) $.
\end{lemma}

\begin{lemma} \label{BCH_formula}
	(from \cite{barrau2018invariant}) The exponential mapping product of two Lie algebras can be obtained from the Baker-Campbell-Hausdorff (BCH) formula in the following form: 
	\begin{small}
		\begin{equation} 
			\begin{aligned}
				\exp(A) \exp(B) &= \exp( A + B + \frac{1}{2!} \left[ A, B \right] + \cdots) \\
				&= \exp( A + B + O(\left \| A \right \| ^ {2}, \left \| B \right \| ^ {2}, \left \| A \right \| \left \| B \right \| )) \label{exponential_mapping_product}
			\end{aligned}
		\end{equation}
	\end{small}
	where $ A,B \in \mathfrak{g} $, $ \left[ A, B \right] := AB-BA $ denotes the Lie brackets of elements $A$ and $B$. When $ B=-A $, $ \exp(A) \exp(B) = I_{d}  $ can be obtained.
\end{lemma}

\begin{definition} \label{left_right_action}
	The left and right actions of a group $G$ on a set $X$ are defined as: 
	\begin{subequations} \label{left_right_action_eq}
		\begin{align}
			\text{left action: } & G \times X \to X, \;\;(a, b, \bullet)_{L} \triangleq a \bullet b \\
			\text{right action: } & X \times G \to X, \;\;(a, b, \bullet)_{R} \triangleq b \bullet a
		\end{align}
	\end{subequations}
	where $a$ and $b$ denote the elements of $G$ and $X$, respectively. 
\end{definition}

\begin{definition} \label{Minkowski_sum_on_group}
	Given two sets $ X $ and $ Y $ on the group $ G $, their left-invariant Minkowski sum on the group is defined as $ (X \bigoplus_{G} Y)_{L} = \left\{ x \bullet y \mid x \in X \subseteq G, y \in Y \subseteq G \right\} $, and their right-invariant Minkowski sum on the group is defined as $ (X \bigoplus_{G} Y)_{R} = \left\{ y \bullet x \mid x \in X \subseteq G, y \in Y \subseteq G \right\} $. 
\end{definition}

\begin{remark} \label{left_right_remark}
	The existence of both left-invariant and right-invariant Minkowski sums on groups is due to the fact that the elements of a group do not necessarily satisfy the exchange law under binary operations $ \bullet $, such as matrix multiplication operations and so on. The group is an Abelian group if all elements of the group satisfy the exchange law under a given binary operations $ \bullet $, when the left-invariant and the right-invariant cases are completely equivalent. For ease of exposition later in this paper, the Minkowski sum on a group can be uniformly denoted as $ (X \bigoplus_{G} Y)_{\varrho} $, which denotes the left-invariant case when $ \varrho = L $ and the right-invariant case when $ \varrho = R  $. Similarly, the left and right action in Definition \ref{left_right_action} are define uniformly as $ c ^ {\varrho} = (a, b, \bullet)_{\varrho} $, where $c ^ {\varrho}$ denotes the variable computed by the left or right action.
\end{remark}

\begin{definition} \label{Zonotope_on_group}
	A zonotope $ \mathbf{X} \subseteq G $ on the group is defined as $ \mathbf{X} = (\mathcal{X} \bigoplus_{G} \mathcal{H} )_{\varrho} $, where the element $ \mathcal{X} \in G $ is the center of $ \mathbf{X} $ and $ \mathcal{H} = \exp(\varXi) = \exp(\left[ \varepsilon \right] ^ {\land}) = \exp( \left[ \left \langle 0, H \right \rangle  \right] ^ {\land} ) $ is the zonotopic set of the error. $ H \in \mathbb{R} ^ {d \times m} $ is the generator matrix. 
\end{definition}

\subsection{Problem Formulation}
\label{subsec:Problem Formulation}
Consider the following discrete system on the group \cite{barrau2018invariant}:
\begin{subequations} \label{sys_eq}
	\begin{align}
		\mathcal{X}_{k+1} &= f(\mathcal{X}_{k}, u_{k}) \bullet \exp([D_{w} w_{k}]^{\land}) \label{sys_eq(a)} \\
		\mathcal{Y}_{k} &= h(\mathcal{X}_{k}) + D_{v} v_{k} \label{sys_eq(b)}
	\end{align}
\end{subequations}
where $\mathcal{X}_{k} \in G$ is the system state variable, $u_{k} \in \mathbb{R}^{n_{u}}$ is the control input variable,  $w_{k} \in \mathbb{R}^{n_{w}}$ is an unknown vector representation of the process disturbances, $\mathcal{Y}_{k} \in \mathbb{R}^{n_{y}}$ is measured output variable, and $v_{k} \in \mathbb{R}^{n_{v}}$ is an unknown vector encoding the measurement noise. $f$ is a nonlinear function satisfying the group affine property \eqref{group_affine_property}. $D_{w}$ and $D_{v}$ are matrices of appropriate dimensions.

\begin{remark} \label{sys_remark}
	Systems satisfying the system model \eqref{sys_eq} are widely available. For a two-frames system \cite{9689974}, the traditional state equation defined in Euclidean space can be transformed into the system model \eqref{sys_eq} on the group. The two-frames refer to the earth(or fixed) and body coordinate systems introduced when describing rigid moveable objects. Such systems are widely existent, e.g., cars, robots, airplanes, etc. It is natural for such the system to have a rotation matrix $ R $ when describing the transformation between the two coordinates. According to the knowledge of Group Theory, it is known that the rotation matrix naturally constitutes a special orthogonal group $ SO(n) $, where $ n $ is the dimension, which can be further combined with the translation vectors to form the special Euclidean group $ SE(n) $ if translational transformations are also considered. Taking the matrix elements in the above group as new state variables, the system model \eqref{sys_eq} can be easily and equivalently obtained, for a more detailed description, please refer to the \cite{9689974}.
\end{remark} 

\begin{remark} \label{discretization_remark}
	It is worth mentioning that the continuous system $ d\mathcal{X}_{t}/dt = f_{u_{t},w_{t}}(\mathcal{X}_{t}) = f(\mathcal{X}_{t}, u_{t}) + \mathcal{X}_{t} w_{t} $ is defined in the \cite{7523335}, where the matrix variable $ \mathcal{X}_{t} $ is naturally embedded in the $ SE(n)$ Lie group space. However, this paper considers discrete systems, for a group dynamical system, the way of discretization affects its group structure, a common discretization method to maintain the group structure is to use an exponential mapping: 
	\begin{subequations}
		\begin{align}
			\text{left-invariant case: }\mathcal{X}_{k+1} = \mathcal{X}_{k} \bullet \exp (\delta \mathcal{T}_{\mathcal{X}_{k}}(f_{u_{k},w_{k}})) \label{sys_eq_Lie_dis_left} \\
			\text{right-invariant case: }\mathcal{X}_{k+1} =  \exp (\delta \mathcal{T}_{\mathcal{X}_{k}}(f_{u_{k},w_{k}})) \bullet \mathcal{X}_{k} \label{sys_eq_Lie_dis_right}
		\end{align}
	\end{subequations}
	where $ \delta $ is the sampling period. The discretized matrix variable $ \mathcal{X}_{k} $ can still maintain the group structure of the original continuous system. If the right-invariant case is considered, the \eqref{sys_eq(a)} can be changed to $ \mathcal{X}_{k+1} = \exp([D_{w} w_{k}]^{\land}) \bullet f(\mathcal{X}_{k}, u_{k}) $. 
\end{remark}

\begin{assumption} \label{Uncertainty_assumption}
	The initial state of the system $ \mathcal{X}_{0} $, the process disturbance $w_{k}$, and the measurement noise $v_{k}$ are all unknown (the statistical properties of the noise or disturbance are not known), but it is known that these variables are bounded as $ \mathcal{X}_{0} \in \mathbf{X}_{0} = (\hat{\mathcal{X}}_{0} \bigoplus_{G} \mathcal{H}_{0} )_{\varrho} $, $ \mathcal{H}_{0} = \exp(\left[ \left \langle 0, H_{0} \right \rangle  \right] ^ {\land}) $, $w_{k} \in W_{k} = \left \langle 0, H_{w}\right \rangle $, and $v_{k} \in V_{k} = \left \langle 0, H_{v}\right \rangle $. $ \hat{\mathcal{X}}_{0} $, $H_{0}$ are the initial estimated state and generator matrix, respectively, which are known. $H_{w}$ and $H_{v}$ are both known generator matrices determined by the upper and lower bounds of the noise.
\end{assumption}

The purpose of this paper is to consider state estimation on Lie groups for a class of group affine systems by using set-membership filtering when the noise are unknown but bounded. Invariant filtering theory is introduced into the set-membership estimation method to extend the application of invariant filtering theory and fill the gap of zonotopic set-membership filtering for state estimation on Lie groups.

\section{Invariant Zonotopic Set-Membership Filtering}
\label{sec:Invariant Zonotopic Set-Membership Filtering}
\subsection{Observations Structure on the Group}
Considering the nonlinear observer on the group \cite{barrau2018invariant}:
\begin{subequations} \label{Observer}
	\begin{align}
		\text{left action case: }\hat{\mathcal{X}}_{k+1} &= f(\hat{\mathcal{X}}_{k}, u_{k}) \bullet \exp( \left[ L_{k}z_{k} \right] ^{\land} ) \label{left_observer_1} \\
		\text{right action case: }\hat{\mathcal{X}}_{k+1} &= \exp( \left[ L_{k}z_{k} \right] ^{\land} ) \bullet f(\hat{\mathcal{X}}_{k}, u_{k}) \label{right_observer_1} \\
		z_{k} &= \mathcal{Y}_{k} - h(\hat{\mathcal{X}}_{k}) \label{measurement_error_equation}
	\end{align}
	where $ L_{k} $ is the observer gain, $ z_{k} \in \mathbb{R}^{n_{y}} $ is called the innovation. 
\end{subequations} 

\begin{definition}
	The invariant estimation error for the two different group action cases can be defined as:
	\begin{subequations} \label{error}
		\begin{align}
			\text{left action case: }\mathcal{\eta}_{k}^{L} &= \hat{\mathcal{X}}_{k}^{-1} \bullet \mathcal{X}_{k} \label{left_error} \\
			\text{right action case: }\mathcal{\eta}_{k}^{R} &= \mathcal{X}_{k} \bullet \hat{\mathcal{X}}_{k}^{-1}.  \label{right_error}
		\end{align}
	\end{subequations}
\end{definition}

\begin{lemma} \label{Measurement_linear_property}
	(from \cite{barrau2018invariant}) In the innovation \eqref{measurement_error_equation}, as $ \epsilon_{k}^{\varrho} $ is assumed to be small, and as $ \exp([0]^{\land}) = I_{d} $, a first-order Taylor expansion in $ \epsilon_{k}^{\varrho} \in \mathbb{R}^{d} $ arbitrary allows defining $ C_{k} ^ { \varrho } $ as  
	\begin{equation} 
		h( (\hat{\mathcal{X}}_{k}, \exp([\epsilon_{k}^{\varrho}]^{\land}), \bullet)_{\varrho} )  - h(\hat{\mathcal{X}}_{k}) := C_{k}^{\varrho} \epsilon_{k}^{\varrho} + O(\left \| \epsilon_{k}^{\varrho} \right \| ^ {2} ). \label{measurement_linear_property_eq} 
	\end{equation}
\end{lemma}

\begin{remark} \label{Measurement_equation_modification}
	It has been shown in the \cite{barrau2018invariant} that for \eqref{measurement_error_equation}, if $ h(\hat{\mathcal{X}}_{k}) $ is represented linearly, i.e., $ h(\hat{\mathcal{X}}_{k}) = \hat{\mathcal{X}}_{k} \bigstar b $, $ b $ is a known vector, and $ \bigstar $ is a operation symbol defined in the \cite{BARRAU201936}, then we can choose an alternative innovation $ z_{k} = \hat{\mathcal{X}}_{k}^{-1} \bigstar \mathcal{Y}_{k} - b $, where the corresponding matrix $ C_{k}^{\varrho} $ will not depend on the state estimation $ \hat{\mathcal{X}}_{k} $ and become a constant matrix $ C^{\varrho} $ \cite{barrau2018invariant}. 
\end{remark}

\begin{theorem} \label{Update_Law}
	For the system \eqref{sys_eq}, $ \mathcal{X}_{k+1} $ can be contained by the zonotope $ (\hat{\mathcal{X}}_{k+1} \bigoplus_{G} \mathcal{H}_{k+1}^{\varrho} )_{\varrho} $ if the state at moment $ k $ satisfies $ \mathcal{X}_{k} \in (\hat{\mathcal{X}}_{k} \bigoplus_{G} \mathcal{H}_{k}^{\varrho} )_{\varrho} $, $ \mathcal{H}_{k}^{\varrho} = \exp( \varXi_{k}^{\varrho} ) = \exp( \left[ \varepsilon_{k}^{\varrho} \right] ^ {\land} ) $, which satisfies
	\begin{subequations}
		\begin{align}
			\hat{\mathcal{X}}_{k+1} \! &= \! (f(\hat{\mathcal{X}}_{k}, u_{k}), \exp( \left[ L_{k}z_{k} \right] ^{\land} ), \bullet )_{\varrho} \label{theorem1_observer} \\
			\mathcal{H}_{k+1}^{\varrho} \! &= \! \exp( \varXi_{k+1}^{\varrho} ) = \exp( \left[ \varepsilon_{k+1}^{\varrho} \right] ^ {\land} ) \label{theorem1_update_error_set_on_group} \\
			\varepsilon_{k+1}^{\varrho} \! &= \! \left \langle 0, H_{k+1}^{\varrho} \right \rangle \label{theorem1_error_set_on_E} \\
			H_{k+1}^{\varrho} \! &= \! \left[ (A_{k}^{\varrho}-L_{k}C_{k}^{\varrho}) \mathcal{R}_{s}(H_{k}^{\varrho}) ; D_{w}H_{w} ; -L_{k}D_{v}H_{v} \right] \label{theorem1_H_update}
		\end{align}
	\end{subequations} 
	where $ A_{k}^{\varrho} $ is defined to satisfy Lemma \ref{Fundamental_property_of_invariant_filtering} and $ C_{k}^{\varrho} $ is defined to satisfy Lemma \ref{Measurement_linear_property}.
\end{theorem}

\begin{IEEEproof}
	For the left-invariant case, $ \varrho = L $, then there are $ \hat{\mathcal{X}}_{k+1} = (f(\hat{\mathcal{X}}_{k}, u_{k}), \exp( \left[ L_{k}z_{k} \right] ^{\land} ), \bullet )_{L} = f(\hat{\mathcal{X}}_{k}, u_{k}) \bullet \exp( \left[ L_{k}z_{k} \right] ^{\land} )  $ and \eqref{sys_eq(a)}. Following the definition of notations in Section \ref{subsec:preliminaries}, let $ \eta_{k+1}^{L} \in \mathcal{H}_{k+1}^{L} $, $ \epsilon_{k+1}^{L} \in \varepsilon_{k+1}^{L} $, and $ \eta_{k+1}^{L} = \exp([\epsilon_{k+1}^{L}]^{ \land }) $. By using $ (\exp(x))^{-1} = \exp(-x) $,  $ [a]^{ \land } + [b]^{ \land } = [a+b]^{ \land }  $, $ \exp([a] ^ { \land }) = I_{d} + [a] ^ { \land } + O ([a] ^ { \land }) $, Lemma \ref{Fundamental_property_of_invariant_filtering}, and Lemma \ref{BCH_formula}, the left-invariant estimation error is calculated as 
	\begin{small}
		\begin{equation*}
			\begin{aligned}
				\mathcal{\eta}_{k+1}^{L} &= \! \hat{\mathcal{X}}_{k+1}^{-1} \bullet \mathcal{X}_{k+1} \\
				&= \! (f(\hat{\mathcal{X}}_{k}, u_{k}) \! \bullet \! \exp( \left[ L_{k}z_{k} \right] ^{\land} ) )^{-1} \! \bullet \! f(\mathcal{X}_{k}, u_{k}) \! \bullet \! \exp( [ D_{w} w_{k} ] ^{\land}) \\
				&= \! \exp([-L_{k}z_{k}]^{\land}) \! \bullet \! f^{-1}(\hat{\mathcal{X}}_{k}, u_{k}) \! \bullet \! f(\mathcal{X}_{k}, u_{k}) \! \bullet \! \exp( [ D_{w} w_{k} ] ^{\land}) \\
				&= \! \exp([-L_{k}z_{k}]^{\land}) \bullet  g(\hat{\mathcal{X}}_{k}^{-1} \bullet \mathcal{X}_{k}, u_{k}) \bullet \exp( [ D_{w} w_{k} ] ^{\land}) \\
				&= \! \exp([-L_{k}z_{k}]^{\land}) \bullet  g(\mathcal{\eta}_{k}^{L}, u_{k}) \bullet \exp( [ D_{w} w_{k} ] ^{\land}) \\
				&= \! \exp([-L_{k}z_{k}]^{\land}) \bullet  g(\exp([\epsilon_{k}^{L}] ^{\land}), u_{k}) \bullet \exp( [ D_{w} w_{k} ] ^{\land}) \\ 
				&\approx \! \exp([-L_{k}z_{k}]^{\land}) \bullet  \exp([A_{k}^{L} \epsilon_{k}^{L}] ^{\land}) \bullet \exp( [ D_{w} w_{k} ] ^{\land}) \\
				&\approx \! (I_{d} + [-L_{k}z_{k}] ^ { \land }) (I_{d} + [A_{k}^{L} \epsilon_{k}^{L}] ^ { \land }) ( I_{d} + [D_{w} w_{k}] ^ { \land } ) \\
				&\approx \! I_{d} + [A_{k}^{L}\epsilon_{k}^{L} + D_{w}w_{k} - L_{k}z_{k}]^{\land} \\
				&\approx \! \exp([A_{k}^{L}\epsilon_{k}^{L} + D_{w}w_{k} - L_{k}z_{k}]^{\land}).
			\end{aligned}
		\end{equation*}
	\end{small}
	The innovation $ z_{k} $ can be calculated as $ z_{k} = \mathcal{Y}_{k} - h(\hat{\mathcal{X}}_{k}) = h(\mathcal{X}_{k}) - h(\hat{\mathcal{X}}_{k}) + D_{v}v_{k} = h(\hat{\mathcal{X}}_{k} \bullet \exp([\epsilon_{k}^{L}] ^{\land})) - h(\hat{\mathcal{X}}_{k}) + D_{v}v_{k} \approx C_{k}^{L} \epsilon_{k}^{L} + D_{v}v_{k} $ by using Lemma \ref{Measurement_linear_property} and $ \mathcal{X}_{k} = \hat{\mathcal{X}}_{k} \bullet \mathcal{\eta}_{k}^{L} = \hat{\mathcal{X}}_{k} \bullet \exp([\epsilon_{k}^{L}] ^{\land}) $.
	Thus, we have $ \exp([\epsilon_{k+1}^{L}]^{ \land }) = \eta_{k+1}^{L} = \exp([(A_{k}^{L}-L_{k}C_{k}^{L}) \epsilon_{k}^{L} + D_{w}w_{k} - L_{k}D_{v}v_{k})]^{\land}) $. The error evolution equation in vector space can be linearized as: 
	\begin{equation}
		\epsilon_{k+1}^{L} = (A_{k}^{L}-L_{k}C_{k}^{L}) \epsilon_{k}^{L} + D_{w}w_{k} - L_{k}D_{v}v_{k}. \label{linear_error_dynamic}
	\end{equation}
	
	Thus, in vector space, the evolution of the error reachable set described by zonotope can be represented as $ \varepsilon_{k+1}^{L} = (A_{k}^{L}-L_{k}C_{k}^{L}) \varepsilon_{k}^{L} \bigoplus D_{w}W_{k} \bigoplus (-L_{k}D_{v})V_{k} $. And since $ \varepsilon_{k+1}^{L} = \left \langle 0, H_{k+1}^{L} \right \rangle $, $ W_{k} = \left \langle 0, H_{w} \right \rangle $, and $ V_{k} = \left \langle 0, H_{v} \right \rangle $, using Property \ref{property_zonotope_linear} and \ref{property_zonotope_reduction}, we can get $ H_{k+1}^{L} = \left[ (A_{k}^{L}-L_{k}C_{k}^{L}) \mathcal{R}_{s}(H_{k}^{L}) ; D_{w}H_{w} ; -L_{k}D_{v}H_{v} \right] $. 
	
	There is a similar proof for the right-invariant case, where $ \hat{\mathcal{X}}_{k+1} = (f(\hat{\mathcal{X}}_{k}, u_{k}), \exp( \left[ L_{k}z_{k} \right] ^{\land} ), \bullet )_{R} =  \exp( \left[ L_{k}z_{k} \right] ^{\land} ) \bullet f(\hat{\mathcal{X}}_{k}, u_{k}) $ and $ \mathcal{X}_{k+1} = \exp([D_{w} w_{k}]^{\land}) \bullet f(\mathcal{X}_{k}, u_{k}) $. 
\end{IEEEproof}

\begin{remark}
	When choosing to apply alternative innovations as described in Remark \ref{Measurement_equation_modification}, the error evolution equation \eqref{linear_error_dynamic} is derived in the same process and form. Moreover, it is further shown in the \cite{barrau2018invariant} that the choice of the application of the above alternative innovations will in fact yield the same estimation results.
\end{remark}

\subsection{Observations Gain Tuning}
\label{subsec:Observations Gain Tuning}
Two methods for designing the observer gain are presented below, namely the pole configuration method and the optimal gain calculation method under the F-radius criterion. For the system \eqref{sys_eq} on the group, the corresponding linear error dynamic system \eqref{linear_error_dynamic} can be obtained under the observer \eqref{Observer}. According to the modern control theory, for a linear observable system, it is generally natural to consider the pole configuration method for observer gain tuning, because this ensures the system stability. Another approach is that when the zonotope is used to describe the uncertainty of the system states, a gain optimization strategy based on the F-radius criterion can be used, which is proposed in the \cite{COMBASTEL2015265}. When performing state estimation, the smaller the uncertainty of the estimated state is, the better, based on the guarantee of accuracy. The second method is to make the uncertainty set of the state estimations as small as possible by minimizing the trace of the zonotope generator matrix during the estimation process to improve the effectiveness of the state estimation.

\subsubsection{Pole Configuration Method}
\begin{theorem} \label{Pole_Configuration}
	For an observable system with the linear error dynamic \eqref{linear_error_dynamic}, the observer gain $ L_{k} $ can be designed by the pole configuration method such that $ A_{k}^{\varrho}-L_{k}C_{k}^{\varrho} = \bar{A} $, where $ \bar{A} $ satisfies spectral radius $ \rho(\bar{A}) < 1 $ and $ w_{k} $, $ v_{k} $ are bounded, which ensures the bounded stability of the error dynamic system, and furthermore the stability of the state estimation of the system \eqref{sys_eq} can be guaranteed. 
\end{theorem}
\begin{IEEEproof}
	By pole configuration, such that $ A_{k}^{\varrho}-L_{k}C_{k}^{\varrho} = \bar{A} $ where $ \bar{A} $ is a constant matrix with all eigenvalues less than 1. Then equation \eqref{linear_error_dynamic} can be written $ \epsilon_{k+1}^{\varrho} = \bar{A} \epsilon_{k}^{\varrho} + D_{w}w_{k} - L_{k}D_{v}v_{k} $. By iteration, $ \epsilon_{k+1}^{\varrho} = \bar{A}^{k+1} \epsilon_{0}^{\varrho} + \begin{matrix} \sum_{i=0}^k \bar{A}^{i} D_{w}w_{k-i} \end{matrix} + \begin{matrix} \sum_{i=0}^k \bar{A}^{i} (- L_{k-i}D_{v}v_{k-i}) \end{matrix}  $ can be obtained. Since $ \rho(\bar{A}) < 1 $, there is $ \lim_{k \to +\infty}\bar{A}^{k} = 0 $. Therefore, $ \bar{A}^{k+1} \epsilon_{0}^{\varrho} \to 0 $ when $ k \to +\infty $. When considering noise, due to the Assumption \ref{Uncertainty_assumption}, there are $ w_{k} $ and $ v_{k} $ that are bounded, and hence the error dynamic system is bounded stable according to Bounded-Input Bounded-Output (BIBO) stability. Since $ \mathcal{\eta}_{k}^{\varrho} = \exp([\epsilon_{k}^{\varrho}] ^{\land}) $, it follows that $ \mathcal{\eta}_{k}^{\varrho} $ is also bounded. Also since $ \mathcal{X}_{k} = \hat{\mathcal{X}}_{k} \bullet \mathcal{\eta}_{k}^{\varrho} $, the state estimation and the true state deviation are bounded, ensuring the stability of the state estimation. 
\end{IEEEproof}

The execution flow of the InZSMF algorithm based on pole configuration method is shown in Algorithm \ref{alg1}.

\begin{algorithm}[htbp]
	\caption{InZSMF Algorithm based on Pole Configuration.}
	\label{alg1}
	\begin{algorithmic}
		\STATE 
		\STATE {\textsc{INITIALIZATION}:} $\hat{\mathcal{X}}_{0}$, $H_{0}^{\varrho}$, $H_{w}$, $H_{v}$, $D_{w}$, $D_{v}$, and pole configuration vector $ \mathbf{\lambda} = [ \lambda_{1}, \dots, \lambda_{n} ] $ .
		\STATE {\textsc{LOOP}}
		\STATE \hspace{0.5cm} $ A_{k}^{\varrho} $ is defined as satisfying $ g(\exp([\epsilon_{k}^{L}] ^{\land}), u_{k}) = \exp([A_{k}^{L} \epsilon_{k}^{L}] ^{\land}) $ and $ C_{k}^{\varrho} $ is defined as satisfying $ h(\hat{\mathcal{X}}_{k} \bullet \exp([\epsilon_{k}^{L}] ^{\land})) - h(\hat{\mathcal{X}}_{k}) = C_{k}^{L} \epsilon_{k}^{L} $. 
		\STATE \hspace{0.5cm} {\textbf{Propagation}}
		\STATE \hspace{1cm} $ \hat{\mathcal{X}}_{k+1|k} = f(\hat{\mathcal{X}}_{k}, u_{k}) $
		\STATE \hspace{0.5cm} {\textbf{Measurement update}}
		\STATE \hspace{1cm} $ z_{k} = \mathcal{Y}_{k} - h(\hat{\mathcal{X}}_{k}) $
		\STATE \hspace{1cm} \textbf{Compute the observer gain}: The observer gain $L_{k}$ \\ \quad \quad \quad \; is computed based on $ A_{k}^{\varrho} $, $C_{k}^{\varrho} $, and $ \mathbf{\lambda} $, by using the \\ \quad \quad \quad \; pole configuration method. 
		\STATE \hspace{1cm} $ \hat{\mathcal{X}}_{k+1} = (\hat{\mathcal{X}}_{k}, \exp([L_{k}z_{k}]^{\land}), \bullet)_{\varrho} $
		\STATE \hspace{1cm} $ \bar{H}_{k}^{\varrho} = \mathcal{R}_{s}(H_{k}^{\varrho}) $
		\STATE \hspace{1cm} $ H_{k+1}^{\varrho} = \left[ (A_{k}^{\varrho}-L_{k}C_{k}^{\varrho}) \bar{H}_{k}^{\varrho} ; D_{w}H_{w} ; -L_{k}D_{v}H_{v} \right] $
		\STATE \hspace{1cm} $ \varepsilon_{k+1}^{\varrho} = \left \langle 0, H_{k+1}^{\varrho} \right \rangle $
		\STATE \hspace{1cm} $ \mathcal{H}_{k+1}^{\varrho} = \exp( \left[ \varepsilon_{k+1}^{\varrho} \right] ^ {\land} ) $
		\STATE \hspace{1cm} $ \mathbf{X}_{k+1} = (\hat{\mathcal{X}}_{k+1} \bigoplus_{G} \mathcal{H}_{k+1}^{\varrho} )_{\varrho}  $
		\STATE {\textsc{END LOOP}}
	\end{algorithmic}
	
\end{algorithm}

\subsubsection{Optimal Observer Gain under the F-radius Criterion}
\begin{theorem} \label{Optimal_Observer_Gain}
	For an observable system with the linear error dynamic \eqref{linear_error_dynamic}, its zonotope is expressed as $ \varepsilon_{k+1}^{\varrho} = (A_{k}^{\varrho}-L_{k}C_{k}^{\varrho}) \varepsilon_{k}^{\varrho} \bigoplus D_{w}W_{k} \bigoplus (-L_{k}D_{v})V_{k} $, where $ \varepsilon_{k+1}^{\varrho} = \left \langle 0, H_{k+1}^{\varrho} \right \rangle $, $ W_{k} = \left \langle 0, H_{w} \right \rangle $, and $ V_{k} = \left \langle 0, H_{v} \right \rangle $, the observer optimal gain $ L_{k}^{*} = argmin( \left \|  H_{k+1}^{\varrho}  \right \|  ^{2} ) $ is given by the following equations:
	\begin{equation}
		\begin{aligned}
			Q_{v} &= D_{v} H_{v} H_{v}^{T} D_{v}^{T} \\
			Q_{w} &= D_{w} H_{w} H_{w}^{T} D_{w}^{T} \\
			P_{k} &= H_{k}^{\varrho} (H_{k} ^ {\varrho}) ^{T} \\
			\bar{P}_{k} &= \mathcal{R}_{s}(H_{k}^{\varrho}) \mathcal{R}_{s}^{T}(H_{k} ^ {\varrho}) \\ 
			S_{k} &= C_{k}^{\varrho} \bar{P}_{k} (C_{k}^{\varrho})^{T} + Q_{v} \\
			L_{k}^{*} &= A_{k}^{\varrho} \bar{P}_{k} (C_{k}^{\varrho})^{T} S_{k}^{-1}.
		\end{aligned}
	\end{equation}
\end{theorem}
\begin{IEEEproof}
	Minimizing the F-radius of $ \varepsilon_{k+1}^{\varrho} = \left \langle 0, H_{k+1}^{\varrho} \right \rangle $ is equivalent to minimizing the trace of the covariance matrix formed by the generator matrix $ J=tr(P_{k+1})=tr(H_{k+1}^{\varrho} (H_{k+1} ^ {\varrho}) ^{T}) = tr((A_{k}^{\varrho}-L_{k}C_{k}^{\varrho}) \bar{P}_{k} (A_{k}^{\varrho}-L_{k}C_{k}^{\varrho})^{T} +  Q_{w} + L_{k}Q_{v}L_{k}^{T}) $. Since $ J $ is convex with respect to $ L_{k} $, $ L_{k}^{*} $ is the value taken at $ \frac{\partial J}{\partial L_{k}} = 0 $. From \eqref{theorem1_H_update} and knowledge related to the derivatives of matrix traces, it follows that: 
	\begin{small}
		\begin{equation*}
			\begin{aligned}
				\frac{\partial J}{\partial L_{k}} &= \!  \frac{\partial tr((A_{k}^{\varrho}-L_{k}C_{k}^{\varrho}) \bar{P}_{k} (A_{k}^{\varrho}-L_{k}C_{k}^{\varrho})^{T} \! + \! Q_{w} \! + \! L_{k}Q_{v}L_{k}^{T})}{\partial L_{k}}  \\
				&= \! -\frac{\partial tr(A_{k}^{\varrho} \bar{P}_{k} (C_{k}^{\varrho})^{T} L_{k}^{T}) }{\partial L_{k}} - \frac{\partial tr(L_{k} C_{k}^{\varrho} \bar{P}_{k} (A_{k}^{\varrho})^{T}) }{\partial L_{k}} \\ & \; \; \; \; + \frac{\partial tr(L_{k} C_{k}^{\varrho} \bar{P}_{k} (C_{k}^{\varrho})^{T} L_{k}^{T}) }{\partial L_{k}} + \frac{\partial tr(L_{k} Q_{v} L_{k}^{T}) }{\partial L_{k}} \\ 
				&= \! - 2 A_{k}^{\varrho} \bar{P}_{k} (C_{k}^{\varrho})^{T} + 2 L_{k} (C_{k}^{\varrho} \bar{P}_{k} (C_{k}^{\varrho})^{T} + Q_{v} ) \\
				L_{k}^{*} &= \! A_{k}^{\varrho} \bar{P}_{k} (C_{k}^{\varrho})^{T} ( C_{k}^{\varrho} \bar{P}_{k} (C_{k}^{\varrho})^{T} + Q_{v} )^{-1} \\
				&= \! A_{k}^{\varrho} \bar{P}_{k} (C_{k}^{\varrho})^{T} S_{k}^{-1}.
			\end{aligned}
		\end{equation*}
	\end{small}
\end{IEEEproof}

The execution flow of the InZSMF algorithm based on optimal observer gain method under the F-radius criterion is shown in Algorithm \ref{alg2}.

\begin{algorithm}[htbp]
	\caption{InZSMF Algorithm based on Optimal Gain.}
	\label{alg2}
	\begin{algorithmic}
		\STATE 
		\STATE {\textsc{INITIALIZATION}:} $\hat{\mathcal{X}}_{0}$, $H_{0}^{\varrho}$, $H_{w}$, $H_{v}$, $D_{w}$, and $D_{v}$.
		\STATE {\textsc{LOOP}}
		\STATE \hspace{0.5cm} $ A_{k}^{\varrho} $ is defined as satisfying $ g(\exp([\epsilon_{k}^{L}] ^{\land}), u_{k}) = \exp([A_{k}^{L} \epsilon_{k}^{L}] ^{\land}) $ and $ C_{k}^{\varrho} $ is defined as satisfying $ h(\hat{\mathcal{X}}_{k} \bullet \exp([\epsilon_{k}^{L}] ^{\land})) - h(\hat{\mathcal{X}}_{k}) = C_{k}^{L} \epsilon_{k}^{L} $. 
		\STATE \hspace{0.5cm} {\textbf{Propagation}}
		\STATE \hspace{1cm} $ \hat{\mathcal{X}}_{k+1|k} = f(\hat{\mathcal{X}}_{k}, u_{k}) $
		\STATE \hspace{0.5cm} {\textbf{Measurement update}}
		\STATE \hspace{1cm} $ z_{k} = \mathcal{Y}_{k} - h(\hat{\mathcal{X}}_{k}) $
		\STATE \hspace{1cm} \textbf{Compute the optimal observer gain}: 
		\STATE \hspace{1.5cm} $ \bar{H}_{k}^{\varrho} = \mathcal{R}_{s}(H_{k}^{\varrho}) $
		\STATE \hspace{1.5cm} $ Q_{v} = D_{v} H_{v} H_{v}^{T} D_{v}^{T} $ 
		\STATE \hspace{1.5cm} $ \bar{P}_{k} = \bar{H}_{k}^{\varrho} (\bar{H}_{k}^{\varrho}) ^{T} $ 
		\STATE \hspace{1.5cm} $ S_{k} = C_{k}^{\varrho} \bar{P}_{k} (C_{k}^{\varrho})^{T} + Q_{v} $ 
		\STATE \hspace{1.5cm} $ L_{k}^{*} = A_{k}^{\varrho} \bar{P}_{k} (C_{k}^{\varrho})^{T} S_{k}^{-1} $ 
		\STATE \hspace{1cm} $ \hat{\mathcal{X}}_{k+1} = (\hat{\mathcal{X}}_{k}, \exp([L_{k}^{*}z_{k}]^{\land}), \bullet)_{\varrho} $
		\STATE \hspace{1cm} $ H_{k+1}^{\varrho} = \left[ (A_{k}^{\varrho}-L_{k}^{*}C_{k}^{\varrho}) \bar{H}_{k}^{\varrho} ; D_{w}H_{w} ; -L_{k}^{*}D_{v}H_{v} \right] $
		\STATE \hspace{1cm} $ \varepsilon_{k+1}^{\varrho} = \left \langle 0, H_{k+1}^{\varrho} \right \rangle $
		\STATE \hspace{1cm} $ \mathcal{H}_{k+1}^{\varrho} = \exp( \left[ \varepsilon_{k+1}^{\varrho} \right] ^ {\land} ) $
		\STATE \hspace{1cm} $ \mathbf{X}_{k+1} = (\hat{\mathcal{X}}_{k+1} \bigoplus_{G} \mathcal{H}_{k+1}^{\varrho} )_{\varrho}  $
		\STATE {\textsc{END LOOP}}
	\end{algorithmic}
\end{algorithm}

\section{Simulation Experiment}
\label{sec:simulation_experiment}
Consider a 2D planar kinematic model of a vehicle. In the earth coordinate frame, its heading angle is ${\theta \in [0, 2\pi]}$ and its plane position vector is $ \mathbf{x} = [x^{(1)} \; x^{(2)}]^\top $. In the vehicle-body coordinate frame, the vehicle's forward speed is $ \nu^{l} $ and its rotational angular velocity is $ \omega $. Consider its process noise as $ w = [w^{\theta} \; w^{l} \; w^{tr}]^\top $, where $ w^{\theta} $ is the differential odometry error, $ w^{l} $ is the longitudinal (vehicle forward direction) odometry error, and $ w^{tr} $ is the vehicle transversal odometry error. The continuous state equation is shown below \cite{7523335}:
\begin{subequations} \label{car_eq_con}
	\begin{align}
		\frac{\mathrm{d}}{\mathrm{d}t}\theta_{t} &= \omega_{t} + w_{t}^{\theta} \label{car_eq_con_1}\\
		\frac{\mathrm{d}}{\mathrm{d}t}x_{t}^{(1)} &= \cos(\theta_{t})(\nu_{t}^{l}+w_{t}^{l}) - \sin(\theta_{t})(w_{t}^{tr}) \\
		\frac{\mathrm{d}}{\mathrm{d}t}x_{t}^{(2)} &= \sin(\theta_{t})(\nu_{t}^{l}+w_{t}^{l}) + \cos(\theta_{t})(w_{t}^{tr}). 
	\end{align}
\end{subequations}
Its system state equation \eqref{car_eq_con} after discretization is shown in \eqref{car_eq} below by using Euler's method, where $ \delta $ is the sampling period and $ k $ is the number of discretization time steps.
\begin{subequations} \label{car_eq}
	\begin{align}
		\theta_{k+1} &= \theta_{k} + (\omega_{k} + w_{k}^{\theta})\delta \label{car_eq_1} \\
		x_{k+1}^{(1)} &= x_{k}^{(1)} + \cos(\theta_{k})(\nu_{k}^{l}+w_{k}^{l})\delta - \sin(\theta_{k})(w_{k}^{tr})\delta \label{car_eq_2}\\
		x_{k+1}^{(2)} &= x_{k}^{(2)} + \sin(\theta_{k})(\nu_{k}^{l}+w_{k}^{l})\delta + \cos(\theta_{k})(w_{k}^{tr})\delta \label{car_eq_3}
	\end{align}
\end{subequations}

The state estimation problem is to estimate the state of the above system \eqref{car_eq} from the measurement information obtained by the sensor, and the measurement equation is: 
\begin{equation} 
	y_{k} = \begin{bmatrix} x_{k}^{(1)} \\ x_{k}^{(2)} \end{bmatrix} + \begin{bmatrix} v_{k}^{(1)} \\ v_{k}^{(2)}  \end{bmatrix} \label{car_meas}
\end{equation}
where $ v_{k}^{(1)} $ and $ v_{k}^{(2)} $ are the measurement noise in the plane coordinates $x_{k}^{(1)}$ and $x_{k}^{(2)}$, respectively. The conventional ZSMF is to define the state variables of the system \eqref{car_eq} directly in the Euclidean vector space by setting $ x_{k} = [\theta_{k} \; x_{k}^{(1)} \; x_{k}^{(2)}]^\top $, $ u_{k} = [\omega_{k} \; \nu_{k}^{l} \; 0]^\top $, $ w_{k} = [w_{k}^{\theta} \; w_{k}^{l} \; w_{k}^{tr}]^\top $, and $ v_{k} = [v_{k}^{(1)} \; v_{k}^{(2)}]^\top $. Thus equations \eqref{car_eq} can be rewritten as $ x_{k+1} = \varphi(x_{k}, u_{k}, w_{k}) $. We use the Taylor expansion to linearize $ \varphi $ at the state estimate $ \hat{x}_{k} $ and ignore higher order infinitesimal terms. Then, the error dynamical system can be obtained: 
\begin{equation} \label{ZSMF_Error}
	\begin{aligned}
		e_{k+1} &= (A_{k}-L_{k}C)e_{k} + D_{wk}w_{k} - L_{k}v_{k} \\
		A_{k} &= \left. \frac{\partial \varphi}{\partial x_{k}} \right| _{x_{k}=\hat{x}_{k}, w_{k}=0} = \begin{bmatrix}
			1 & 0 & 0\\
			-\sin(\hat{\theta}_{k})\nu_{k}^{l}\delta & 1 & 0 \\ 
			\cos(\hat{\theta}_{k})\nu_{k}^{l}\delta & 0 & 1
		\end{bmatrix} \\
		D_{wk} &= \left. \frac{\partial \varphi}{\partial w_{k}} \right| _{x_{k}=\hat{x}_{k}} = \begin{bmatrix}
			\delta & 0 & 0\\
			0 & \cos(\hat{\theta}_{k})\delta & -\sin(\hat{\theta}_{k})\delta \\ 
			0 & \sin(\hat{\theta}_{k})\delta & \cos(\hat{\theta}_{k})\delta
		\end{bmatrix} \\
		C &= \begin{bmatrix}
			0 & 1 & 0\\
			0 & 0 & 1 
		\end{bmatrix}. 
	\end{aligned}
\end{equation}

The InZSMF method proposed in this paper is considered below. By the way, for the convenience of description, the left-invariant InZSMF (L-InZSMF) is shown as an example of the simulation experiments. It is worth mentioning that, considering the left-invariant case under Remark \ref{Measurement_equation_modification}, the linear evolution of the error dynamics without noise is fully autonomous, which means that the error does not depend on the state trajectory, which is a very good property that facilitates state estimation \cite{7523335}. 

For \eqref{car_eq_1}, there exists an equivalent equation which is written as 
\begin{subequations} \label{rotation}
	\begin{align}
		\frac{\mathrm{d}}{\mathrm{d}t}R(\theta_{t}) &= R(\theta_{t})\exp(\delta[\omega_{t} + w_{t}^{\theta}]_{\times}) \label{rotation_eq} \\
		R(\theta_{t}) &= \begin{bmatrix}
			\cos(\theta_{t}) & -\sin(\theta_{t})  \\
			\sin(\theta_{t}) &  \cos(\theta_{t})
		\end{bmatrix} \label{rotation_matrix_eq2}
	\end{align}
\end{subequations}
where $ R(\theta_{t}) $ denotes the rotation matrix and $ [\centerdot]_{\times} $ denotes the skew-symmetric matrix generation operation (See the Appendix A). 

From \eqref{car_eq_con} and \eqref{rotation_eq}, the state equation on the group after left-invariant discretization is obtained (See the Appendix B):
\begin{equation} \label{Lie_matrix}
	\begin{aligned}
		\mathcal{X}_{k+1} &= \mathcal{X}_{k} \exp(\delta \mathcal{U}_{k}) \exp(\delta \mathcal{W}_{k}) \\
		&= f(\mathcal{X}_{k}, u_k) \exp(\delta [w_{k}]^{\land}) \\
		\mathcal{X}_{k} &= \begin{bmatrix}
			R(\theta_{k}) & \mathbf{x}_{k}  \\
			0_{1\times2} &  1
		\end{bmatrix} \\
		\mathcal{U}_{k} &= \begin{bmatrix}
			[\omega_{k}]_{\times} & \nu_{k}  \\
			0_{1\times2} &  0
		\end{bmatrix} \triangleq [u_{k}]^{\land} \\
		\mathcal{W}_{k} &= \begin{bmatrix}
			[w_{k}^{\theta}]_{\times} & \mathbf{w}_{k}  \\
			0_{1\times2} &  0
		\end{bmatrix} \triangleq [w_{k}]^{\land}
	\end{aligned}
\end{equation}
where $ \mathbf{x}_{k} = [x_{k}^{(1)} \; x_{k}^{(2)}]^{\top} $, $ \nu_{k} = [\nu_{k}^{l} \; 0]^{\top} $, $ \mathbf{w}_{k} = [w_{k}^{l} \; w_{k}^{tr}]^{\top} $, $ u_{k} = [\omega_{k} \; \nu_{k}^{l} \; 0]^{\top} $, and $ w_{k} = [w_{k}^{\theta} \; w_{k}^{l} \; w_{k}^{tr}]^{\top} $. It is easy to show that the state variable $ \mathcal{X}_{k} $ is a discretization variable satisfying the group structure that preserves the Lie group $ SE(2) $, the group multiplication $ \bullet $ is matrix multiplication, and that both $ \mathcal{U}_{k} $ and $ \mathcal{W}_{k} $ are elements of the Lie algebra space of the $ SE(2) $ at the identity element.

The measurement equation \eqref{car_meas} can be transformed as:
\begin{equation} 
	\mathcal{Y}_{k} = \mathcal{X}_{k}\begin{bmatrix} 0_{2\times1}  \\ 1 \end{bmatrix} + \begin{bmatrix} v_{k} \\ 0 \end{bmatrix} \label{car_meas_lie}
\end{equation}
where $ v_{k} = [v_{k}^{(1)}, v_{k}^{(2)}]^{\top} $ is measured noise. 

Considering the left-invariant observation \eqref{left_observer_1} and according to Remark \ref{Measurement_equation_modification} and Theorem \ref{Update_Law}, the linear error dynamic is
\begin{equation} \label{InZSMF_Error}
	\begin{aligned}
		\epsilon_{k+1} &= (A_{k}^{L}-\tilde{L}_{k}\tilde{C}^{L})\epsilon_{k} + \delta w_{k} - \tilde{L}_{k} R(\hat{\theta}_{k})^{\top} v_{k} \\
		A_{k}^{L} &= I_{3} - \delta ad_{u_{k}} \;\;\;\;\;\;
		L_{k} = \tilde{L}_{k} \tilde{p} \\
		\tilde{C}^{L} &= \begin{bmatrix}
			0 & 1 & 0\\
			0 & 0 & 1 
		\end{bmatrix} \;\;\;\;\;\;
		\tilde{p} = \begin{bmatrix}
			1 & 0 & 0\\
			0 & 1 & 0 
		\end{bmatrix} \\
		ad_{u_{k}} &= \begin{bmatrix}
			0 & 0 & 0\\
			0 & 0 & -\omega_{k}\\
			-\nu_{k}^{l} & \omega_{k} & 0
		\end{bmatrix}
	\end{aligned}
\end{equation}
where $ ad_{u_{k}} \in \mathbb{R}^{dim\mathfrak{g} \times dim\mathfrak{g}} $ is the adjoint matrix which satisfies $ ad_{u_{k}}\epsilon_{k} = ad_{\epsilon_{k}}u_{k} $ and $ [\epsilon_{k}]^{\land} [u_{k}]^{\land} - [u_{k}]^{\land} [\epsilon_{k}]^{\land} = [-ad_{u_{k}}\epsilon_{k}]^{\land} $. Thus the generator matrix update law $ H_{k+1}^{L} = \left[ (A_{k}^{L}-\tilde{L}_{k}\tilde{C}^{L}) \mathcal{R}_{s}(H_{k}^{L}) ; \delta H_{w} ; -\tilde{L}_{k} R(\hat{\theta}_{k})^{\top} H_{v} \right] $ for the InZSMF method can be obtained. See the Appendix C for a detailed derivation. 

It is assumed that the vehicle is traveling on a circular ramp with a turning radius $ r=20m $, the traveling speed is constant at $ v=8m/s $, the angular velocity is constant at $ w=v/r=0.4rad/s $, and the sampling period is $ \delta = 0.01s $. The process noise $ w_{k} $ of the system is related to the measurement accuracy of the on-board sensors, and the measurement noise $ v_{k} $ is related to the positioning accuracy of the GPS, assuming that $ w_{k} \in W_{k} = \left \langle 0, diag([0.1, 0.1, 0.1]) \right\rangle $, $ v_{k} \in V_{k} = \left \langle 0, diag([1, 1]) \right\rangle $. In this experiment, the real initial state of the vehicle is assumed to be $ \theta_{0}=\pi/2 $, $ x_{0}^{(1)} = 0 $, $ x_{0}^{(2)} = 0 $. When the order of the generator matrix exceeds $30$, the order reduction operation described in Property \ref{property_zonotope_reduction} is performed.

In the experimental environment where the operation system is Windows 10, the CPU is Intel Core i7-10700F, the RAM is 16G, and the development language is Python 3.9, the comparison experiments are conducted by using the traditional ZSMF and the InZSMF proposed in this paper. The RMSE is used as a metric to evaluate the accuracy of the state estimation, where $ RMSE(\theta) = \sqrt{\frac{1}{N} \begin{matrix} \sum_{k=1}^N ( \theta_{k} - \hat{\theta}_{k} )^2 \end{matrix} } $ and $ RMSE(\mathbf{x}) = \sqrt{\frac{1}{N} \begin{matrix} \sum_{k=1}^N \left[ ( x_{k}^{(1)} - \hat{x}_{k}^{(1)} )^2 + ( x_{k}^{(2)} - \hat{x}_{k}^{(2)} )^2 \right] \end{matrix} } $ measure the accuracy of the angle and position vector state estimation, respectively. The average area ratio (AAR) \cite{10685107} is used as a metric to assess the interval area of the state estimation, where $ AAR(\theta) = \frac{1}{N} \begin{matrix} \sum_{k=1}^N ( \overline{\theta}_{k} - \underline{\theta}_{k} ) \end{matrix} $ and $ AAR(\mathbf{x}) = \frac{1}{N} \begin{matrix} \sum_{k=1}^N ( \overline{x}_{k}^{(1)} - \underline{x}_{k}^{(1)} )( \overline{x}_{k}^{(2)} - \underline{x}_{k}^{(2)} ) \end{matrix} $ measure the interval area of the angle and position vector estimation, respectively. The algorithm average running time (ART) \cite{TANG2022110580} measures the computational time consumption and its unit is seconds in this paper. By the way, in order to reduce the chance of the experiments, each group of experiments is conducted with five independent repetitions, and the average results of the repetitive experiments are shown in Table \ref{tab:ZSMFvsInZSMF} and Table \ref{tab:PCvsFOG}. Among them, the former table shows the experimental results obtained by the InZSMF and ZSMF methods based on F-radius optimal gain under different initial estimated values, and the latter shows the experimental results of different observer gains on state estimation in Algorithm \ref{alg1} and Algorithm \ref{alg2}. In the above tables, the percentage improvement in the performance metrics of InZSMF with respect to the conventional ZSMF method is calculated by using $ -(M_{1}-M_{2})/M_{1} $, where $ M_{1} $ and $ M_{2} $ denote the performance metrics obtained under the ZSMF and InZSMF methods, respectively. Note that all the above three metrics are as small as possible. 

\begin{figure}[htbp]
	\centering
	\subfigure[]{
		\includegraphics[width=0.46\linewidth]{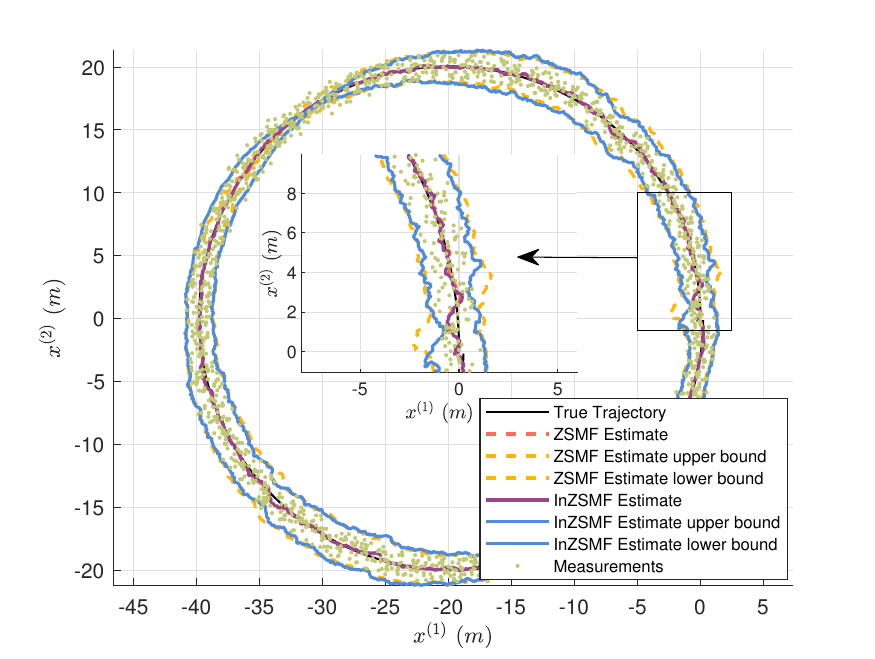}
		\label{INIT_90_0_0:sub1}
	}
	\subfigure[]{
		\includegraphics[width=0.46\linewidth]{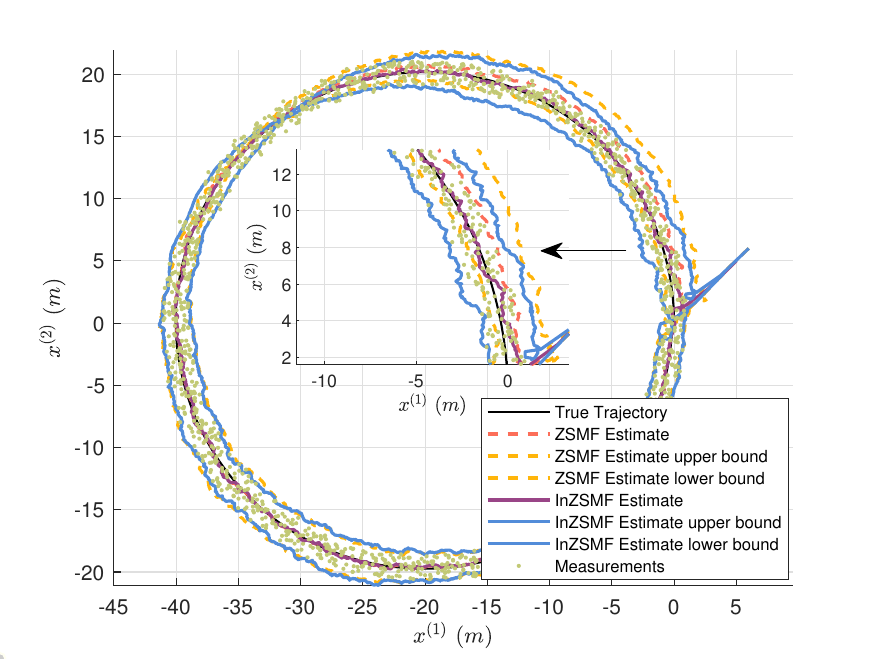}
		\label{INIT_0_5_5:sub1}
	}
	\subfigure[]{
		\includegraphics[width=0.46\linewidth]{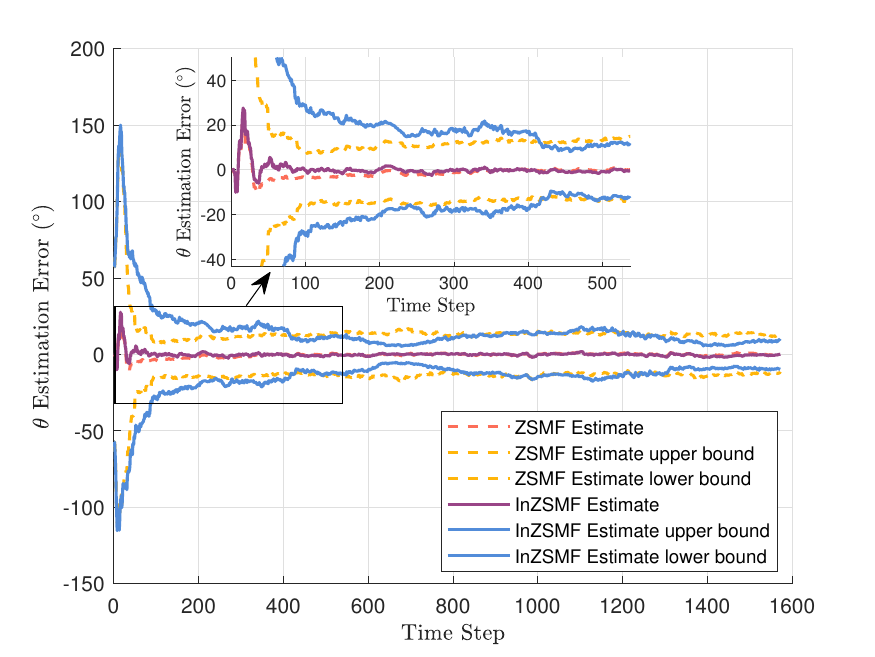}
		\label{INIT_90_0_0:sub2}
	}
	\subfigure[]{
		\includegraphics[width=0.46\linewidth]{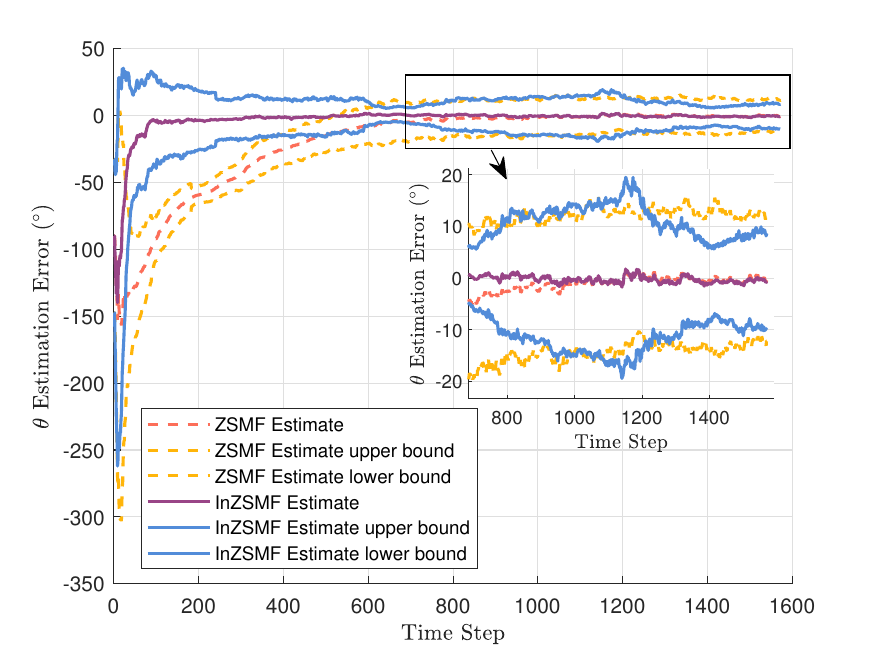}
		\label{INIT_0_5_5:sub2}
	}
	\subfigure[]{
		\includegraphics[width=0.46\linewidth]{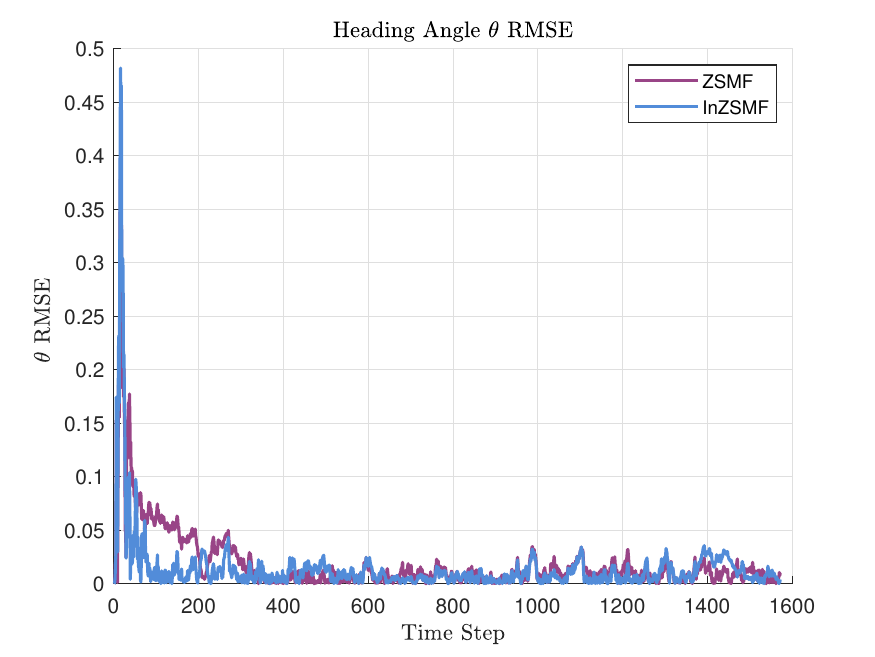}
		\label{INIT_90_0_0:sub3}
	}
	\subfigure[]{
		\includegraphics[width=0.46\linewidth]{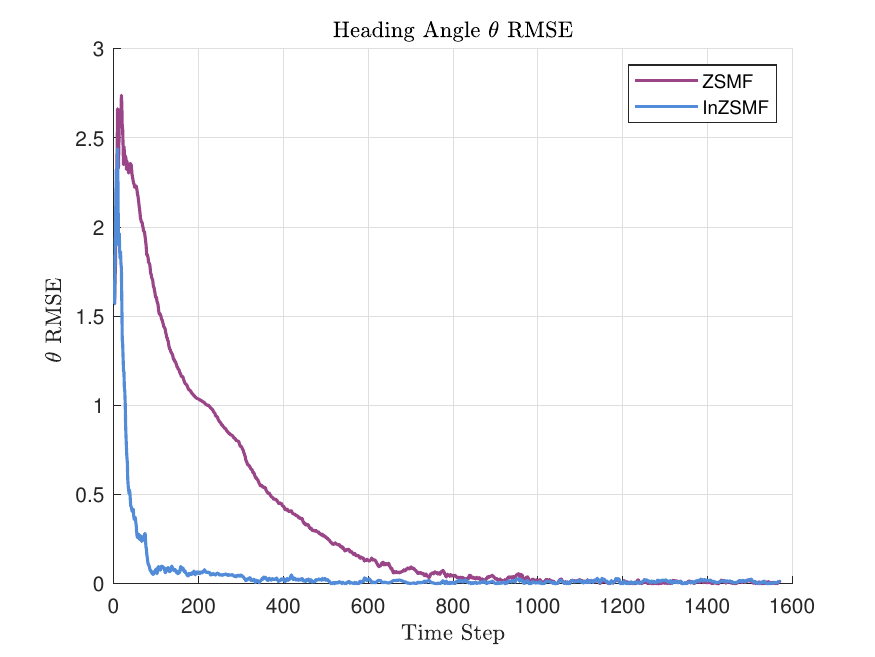}
		\label{INIT_0_5_5:sub3}
	}
	\subfigure[]{
		\includegraphics[width=0.46\linewidth]{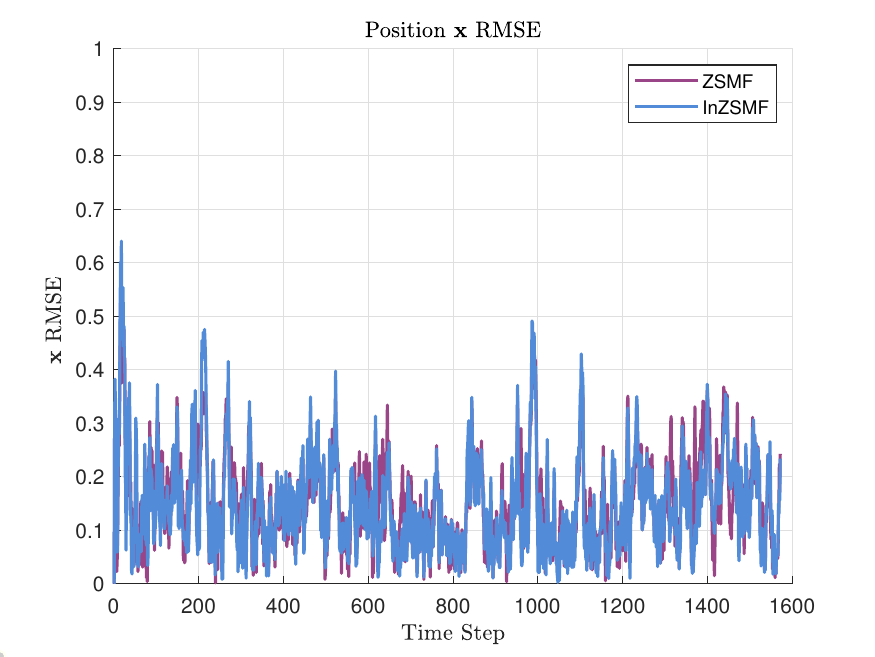}
		\label{INIT_90_0_0:sub4}
	}
	\subfigure[]{
		\includegraphics[width=0.46\linewidth]{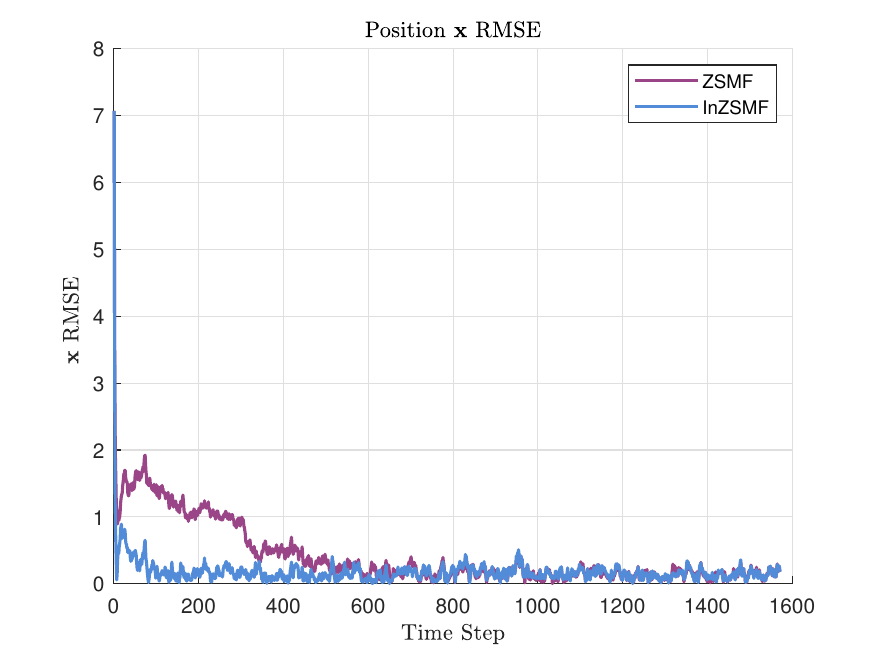}
		\label{INIT_0_5_5:sub4}
	}
	\caption{The comparison of the ZSMF and InZSMF methods. The left column shows the comparison when the initial estimate is $ \hat{\theta}_{0} = \pi / 2 $, $ \hat{x}_{0}^{(1)} = 0 $, $ \hat{x}_{0}^{(2)} = 0 $, and the right column shows the comparison when the initial estimate is $ \hat{\theta}_{0} = 0 $, $ \hat{x}_{0}^{(1)} = 5 $, $ \hat{x}_{0}^{(2)} = 5 $. (a) and (b) show the comparison between the true and estimated trajectories of the vehicle. (c) and (d) show the center estimation error of the vehicle heading angle $\theta$ and its upper and lower bounds. (e) and (f) denote the RMSE metrics for the vehicle heading angle $\theta$. (g) and (h) denote the RMSE metrics for the position vector $ \mathbf{x} $. It can be noticed that when the initial estimation error is small, the behaviors of ZSMF and InZSMF is similar. However, when the initial estimation error is large, the behaviors of the two methods are significantly different. InZSMF has better estimation performance and can rapidly reduce the state estimation error. }
	\label{fig:INIT_diff}
\end{figure}

\subsubsection{Experiment 1 (Comparative experiments of the ZSMF and InZSMF methods) }
In order to validate the effectiveness of the method proposed in this paper under different initial errors, eight sets of comparison experiments are designed to compare the InZSMF method with the traditional ZSMF method. From Table \ref{tab:ZSMFvsInZSMF}, it can be seen that the InZSMF method outperforms the ZSMF method in both RMSE and AAR metrics under different initial estimation errors. About the RMSE performance, especially in the challenging situation where the initial angle estimation $ \hat{\theta_{0}} = 0^{ \circ } $ differ by $ 90^{ \circ } $ and the initial position estimation $ (\hat{x}_{0}^{(1)}, \hat{x}_{0}^{(2)}) = (5, 5) $ differs by $ 5m $, the RMSE of heading angle center estimation can be improved by about $ 53.36\% $ and the RMSE of position vector center estimation can be improved by about $ 50.30\% $ by using the InZSMF method. In addition, from Fig. \ref{fig:INIT_diff}, it can be seen that the convergence speed of the InZSMF method in the state estimation process is significantly improved compared with the ZSMF method, and also the greater the difference in the initial value estimation, the more obvious the improvement effect is. In practice, it is not easy to accurately obtain the initial state of the system \cite{CONG2025111993}. Usually, only a rough initial estimate can be obtained. At this time, the advantages of the InZSMF method are manifested. Regarding the AAR performance, it can be found that the AAR metrics of the heading angle and position vector obtained by the InZSMF method are significantly improved over the ZSMF method, and sometimes the improvement is even more than $ 10\% $. 

\subsubsection{Experiment 2 (Comparative experiments of two observer gain tuning methods) }
\begin{figure}[htbp]
	\centering
	\subfigure[]{
		\includegraphics[width=0.46\linewidth]{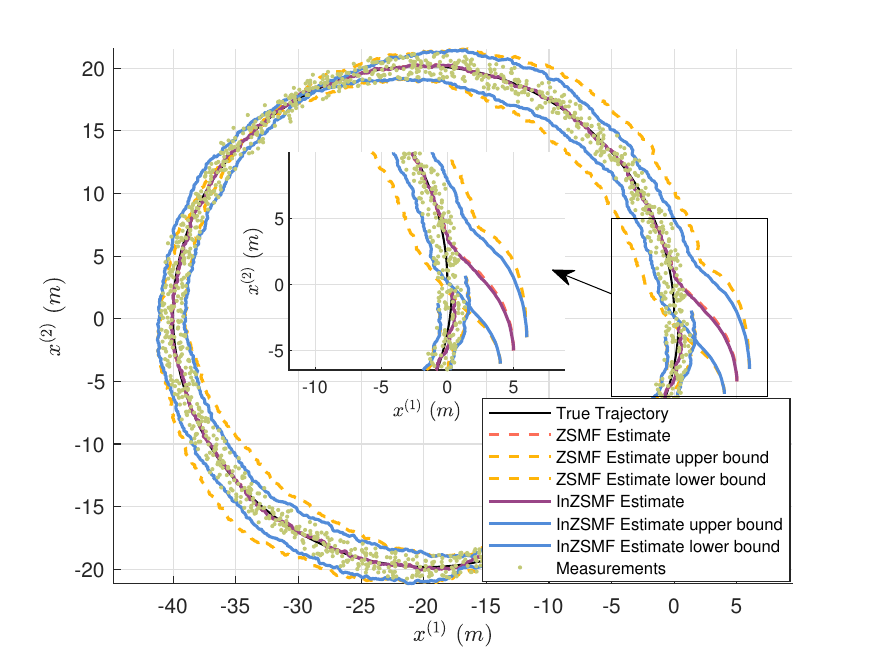}
		\label{AC:sub1}
	}
	\subfigure[]{
		\includegraphics[width=0.46\linewidth]{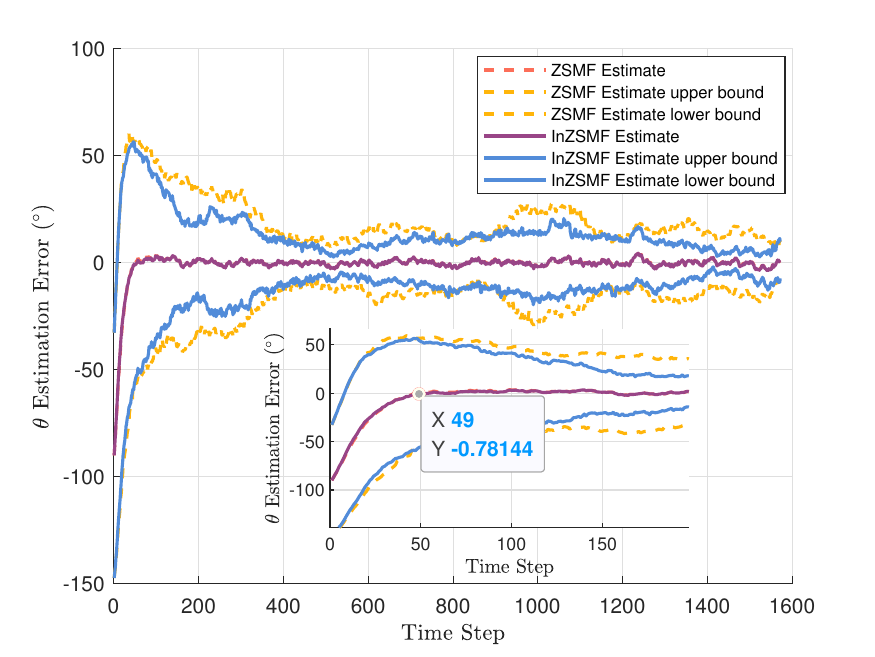}
		\label{AC:sub2}
	}
	\subfigure[]{
		\includegraphics[width=0.46\linewidth]{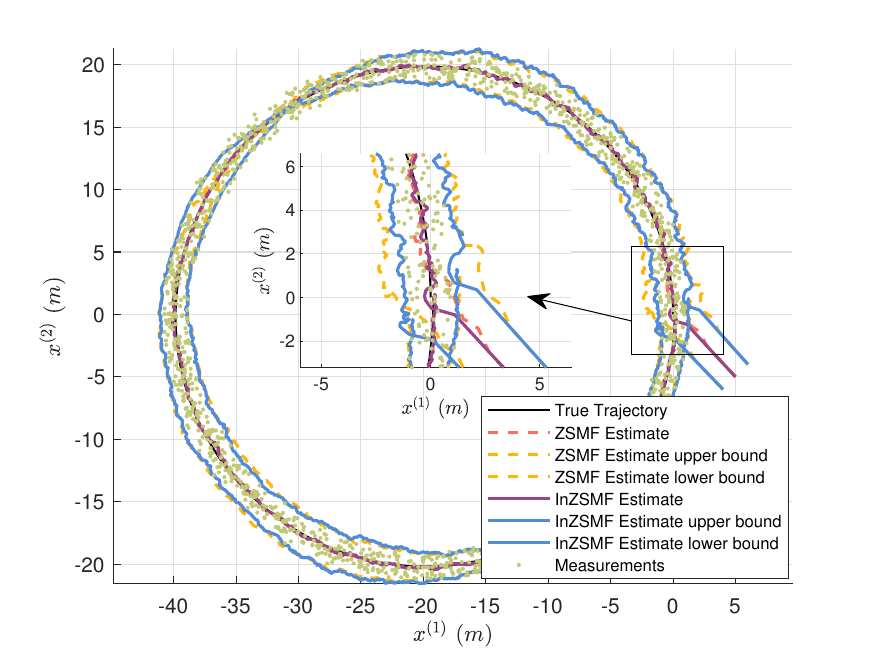}
		\label{AC:sub3}
	}
	\subfigure[]{
		\includegraphics[width=0.46\linewidth]{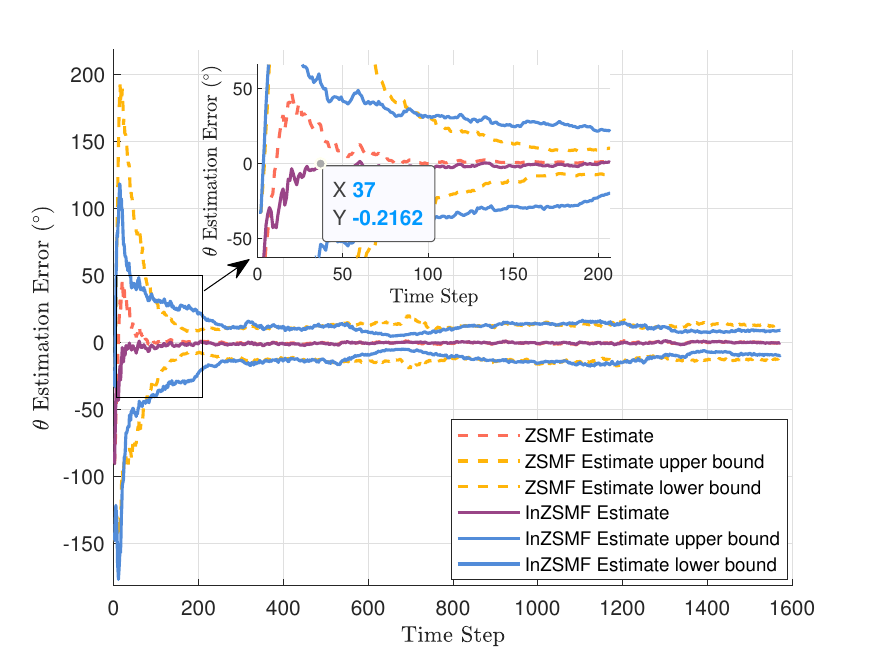}
		\label{AC:sub4}
	}
	\caption{The comparison of the ZSMF and InZSMF methods under two observer gain tuning methods. (a) and (b) respectively present the comparison of the trajectory estimation and the vehicle heading angle estimation errors under the algorithm \ref{alg1} method. (c) and (d) respectively present the comparison of the trajectory estimation and the vehicle heading angle estimation errors under the algorithm \ref{alg2} method. }
	\label{Algorithm Comparison}
\end{figure}

The following analyzes the influence of the two observer gain tuning methods proposed in this paper on the InZSMF and ZSMF methods. The initial states of this experiments are estimated as $ \hat{\theta}_{0} = 0 $, $ \hat{x}_{0}^{(1)} = 5 $, $ \hat{x}_{0}^{(2)} = -5 $. The pole configuration method assumes that the poles are configured as $ [0.95, 0.98, 0.98] $. From Table \ref{tab:PCvsFOG}, it can be seen that the InZSMF method based on Algorithm \ref{alg1} still has improvements in both angle estimation and position vector estimation compared to the ZSMF method, whether it is the RMSE of the zonotope center estimation and the area of the interval estimation corresponding to the zonotope. Especially, the $AAR$ metrics are improved significantly, in which $AAR(\mathbf{x})$ is even improved by $49.94\%$. Further comparison of Algorithm \ref{alg2} under the same initial conditions, it can be found that Algorithm \ref{alg2} is significantly better than Algorithm \ref{alg1} in terms of RMSE performance metrics, whereas Algorithm \ref{alg1} is slightly better than Algorithm \ref{alg2} in terms of the AAR performance metrics. However, the time consumed to run Algorithm \ref{alg2} is much less than that of Algorithm \ref{alg1}. It can be seen in Fig. \ref{Algorithm Comparison} that the convergence speed of the state estimation by using Algorithm \ref{alg2} is better than that of Algorithm \ref{alg1}, but in the case of overshooting, Algorithm \ref{alg1} has more advantages. In conclusion, the InZSMF algorithm can handle the state estimation problem by choosing the appropriate observer gain tuning according to the demand. 

Finally, the computational time consumption of the InZSMF method proposed in this paper is discussed. From the ART metrics in Table \ref{tab:ZSMFvsInZSMF} and \ref{tab:PCvsFOG}, it can be found that the ARTs of the InZSMF and ZSMF methods are in the millisecond range, although the ART of the InZSMF method is slightly increased compared to the traditional method. Algorithm \ref{alg1} proposed in this paper takes no more than $ 5ms $ for a single operation, and Algorithm \ref{alg2} takes no more than $ 1ms $ for a single operation. This indicates that the InZSMF method proposed in this paper, whether based on Algorithm \ref{alg1} or Algorithm \ref{alg2}, has practical application value. 

\begin{table*}[htbp]
	\centering
	\caption{Comparison between the ZSMF and InZSMF Methods based on the F-radius Optimal Gain}
	\begin{tabular}{ccccccc}
		\hline
		Initial estimated value & Methods & ART   & RMSE($\theta$) & RMSE($ \mathbf{x} $) & AAR($\theta$) & AAR($ \mathbf{x} $) \\
		\hline
		$ \hat{\theta}_{0} = \pi / 2 $, $ \hat{x}_{0}^{(1)} = 0 $, $ \hat{x}_{0}^{(2)} = 0 $ & ZSMF  & \textbf{0.0002948} & 0.0297466 & 0.1765084 & 0.5452694 & 4.901758 \\
		& InZSMF & 0.0006729 & \textbf{0.0286901} & \textbf{0.1731275} & \textbf{0.5081545} & \textbf{4.3448098} \\
		&       &       & 3.55\% & 1.92\% & 6.81\% & 11.36\% \\
		$ \hat{\theta}_{0} = \pi / 4 $, $ \hat{x}_{0}^{(1)} = 0 $, $ \hat{x}_{0}^{(2)} = 0 $ & ZSMF  & \textbf{0.0002847} & 0.0607859 & 0.1780535 & 0.5742719 & 4.9474083 \\
		& InZSMF & 0.0005746 & \textbf{0.0588446} & \textbf{0.1756499} & \textbf{0.5264049} & \textbf{4.4215509} \\
		&       &       & 3.19\% & 1.35\% & 8.34\% & 10.63\% \\
		$ \hat{\theta}_{0} = 0 $, $ \hat{x}_{0}^{(1)} = 0 $, $ \hat{x}_{0}^{(2)} = 0 $ & ZSMF  & \textbf{0.0003619} & 0.1388153 & 0.1834197 & 0.5713494 & 4.9350339 \\
		& InZSMF & 0.000676 & \textbf{0.1286731} & \textbf{0.1726175} & \textbf{0.5384568} & \textbf{4.3907049} \\
		&       &       & 7.31\% & 5.89\% & 5.76\% & 11.03\% \\
		$ \hat{\theta}_{0} = \pi / 2 $, $ \hat{x}_{0}^{(1)} = 5 $, $ \hat{x}_{0}^{(2)} = 5 $ & ZSMF  & \textbf{0.0004062} & 0.1430637 & 0.2290812 & 0.590817 & 5.0174807 \\
		& InZSMF & 0.0006746 & \textbf{0.0727635} & \textbf{0.2030627} & \textbf{0.5288208} & \textbf{4.2914036} \\
		&       &       & 49.14\% & 11.36\% & 10.49\% & \textit{\textbf{14.47\%}} \\
		$ \hat{\theta}_{0} = \pi / 4 $, $ \hat{x}_{0}^{(1)} = 5 $, $ \hat{x}_{0}^{(2)} = 5 $ & ZSMF  & \textbf{0.0004163} & 0.0610256 & 0.236809 & 0.5699872 & 4.9573794 \\
		& InZSMF & 0.0006499 & \textbf{0.0538042} & \textbf{0.1980891} & \textbf{0.5102441} & \textbf{4.3881168} \\
		&       &       & 11.83\% & 16.35\% & 10.48\% & 11.48\% \\
		$ \hat{\theta}_{0} = 0 $, $ \hat{x}_{0}^{(1)} = 5 $, $ \hat{x}_{0}^{(2)} = 5 $ & ZSMF  & \textbf{0.0003365} & 0.4768847 & 0.4297928 & 0.5609619 & 4.9347009 \\
		& InZSMF & 0.0005168 & \textbf{0.2223981} & \textbf{0.2135912} & \textbf{0.5469439} & \textbf{4.41341} \\
		&       &       & \textit{\textbf{53.36\%}} & \textit{\textbf{50.30\%}} & 2.50\% & 10.56\% \\
		$ \hat{\theta}_{0} = 0 $, $ \hat{x}_{0}^{(1)} = 5 $, $ \hat{x}_{0}^{(2)} = -5 $ & ZSMF  & \textbf{0.0003384} & 0.1058326 & 0.2370771 & 0.5934819 & 4.9847943 \\
		& InZSMF & 0.0006246 & \textbf{0.0793563} & \textbf{0.2068291} & \textbf{0.5292185} & \textbf{4.3878886} \\
		&       &       & 25.02\% & 12.76\% & \textit{\textbf{10.83\%}} & 11.97\% \\
		$ \hat{\theta}_{0} = 0 $, $ \hat{x}_{0}^{(1)} = -5 $, $ \hat{x}_{0}^{(2)} = 5 $ & ZSMF  & \textbf{0.0004104} & 0.2957031 & 0.2724086 & 0.5594795 & 4.9255123 \\
		& InZSMF & 0.0005910 & \textbf{0.1605700} & \textbf{0.1993127} & \textbf{0.5510282} & \textbf{4.4242021} \\
		&       &       & 45.70\% & 26.83\% & 1.51\% & 10.18\% \\
		\hline
	\end{tabular}%
	\label{tab:ZSMFvsInZSMF}%
\end{table*}%

\begin{table*}[htbp]
	\centering
	\caption{Comparison Method of Pole Configuration and F-radius Optimal Gain}
	\begin{tabular}{rccccc}
		\hline
		\multicolumn{1}{c}{Methods} & ART   & RMSE($\theta$) & RMSE($ \mathbf{x} $) & AAR($\theta$) & AAR($ \mathbf{x} $) \\
		\hline
		\multicolumn{1}{c}{ZSMF(Pole Configuration)} & 0.0029028 & 0.1201428 & 0.5982256 & 0.7416984 & 7.2063406 \\
		\multicolumn{1}{c}{InZSMF(Pole Configuration)} & 0.0030225 & 0.1176755 & 0.5636406 & \textbf{0.5232929} & \textbf{3.6078526} \\
		&       & 2.05\% & 5.78\% & \textit{\textbf{29.45\%}} & \textit{\textbf{49.94\%}} \\
		\multicolumn{1}{c}{ZSMF(F-radius Optimal Gain)} & \textbf{0.0003384} & 0.1058326 & 0.2159713 & 0.5934819 & 4.9847943 \\
		\multicolumn{1}{c}{InZSMF(F-radius Optimal Gain)} & 0.0006246 & \textbf{0.0793563} & \textbf{0.182269} & 0.5292185 & 4.3878886 \\
		&       & \textit{\textbf{25.02\%}} & \textit{\textbf{15.61\%}} & 10.83\% & 11.97\% \\
		\hline
	\end{tabular}%
	\label{tab:PCvsFOG}%
\end{table*}%

\section{Conclusion}
\label{sec:Conclusion}
In this paper, we successfully extend the invariant filtering theory to the field of set-membership filtering and propose the InZSMF method for dealing with the problem of state estimation of unknown but bounded uncertainty invariant discrete systems defined by the zonotope on the group. Through the simulation experiments, it is shown that the InZSMF method is generally superior to the traditional ZSMF method, particularly in some challenging cases. More excitingly, the InZSMF method is capable of completing a single operation in milliseconds of computation time, with the potential to deal with state estimation problems in real time. Regarding the obvious improvement of InZSMF compared with the ZSMF method, we speculate that it may be due to the potential constraints added when transforming the traditional state vectors into state variables on the group, such as the symmetry condition, which makes the state estimation more accurate. This requires further research and discussion in the future. Of course, there are some limitations to the method, which is used on the premise that it must be ensured that the original system state equations can be transformed into group affine systems on the group, and more systems need to be discussed in the future on the basis of two-frames systems such as robots. Overall, the set-membership filtering algorithm optimized by invariant filtering theory has a satisfactory performance improvement, which is similar to the success of invariant filtering in the field of statistical filtering, indicating that invariant filtering theory has a strong research value and application value. 

\appendices
\section*{Appendix}
\subsection*{A. Special Euclidean SE(2)}
$SE(2)$ is the Special Euclidean group used in robotics to represent poses, including translations and rotations, which can be defined as the set of transformation metrics $ G=SE(2):= \left\{ \begin{bmatrix} R(\theta) & \mathbf{x}  \\ 0_{1\times2} &  1 \end{bmatrix}, \theta \in \mathbb{R}, \mathbf{x} = \begin{bmatrix}x_{1} \\ x_{2}\end{bmatrix} \in \mathbb{R}^{2} \right\} $, where $ R(\theta) $ is the rotation matrix defined by \eqref{rotation_matrix_eq2}. The Lie algebra of $SE(2)$ can be expressed as $ \mathfrak{g} = \mathfrak{se}(2) = \left\{ \begin{bmatrix} [\sigma]_{\times} & \mathbf{u}  \\ 0_{1\times2} &  0 \end{bmatrix}; [\sigma]_{\times} = \begin{bmatrix} 0 & -\sigma  \\ \alpha &  0 \end{bmatrix}, \mathbf{u} = \begin{bmatrix}u_{1} \\ u_{2}\end{bmatrix} \in \mathbb{R}^{2} \right\} $. Let $ \boldsymbol{ \zeta } = \begin{bmatrix} \sigma \\ u_{1} \\ u_{2}\end{bmatrix} \in \mathbb{R}^{3} $, then $ [ \boldsymbol{ \zeta } ] ^ { \land } = \begin{bmatrix} [\sigma]_{\times} & \mathbf{u}  \\ 0_{1\times2} &  0 \end{bmatrix} $, and adjoint matrix is $ ad_{\boldsymbol{ \zeta }} = \begin{bmatrix} 0 & 0 & 0 \\ u_{2} & 0 & -\sigma \\ -u_{1} & \sigma & 0 \end{bmatrix} $.  

\subsection*{B. Proof of vehicle state equation on the group}
If the matrix state variable $ \mathcal{X}_{t} = \begin{bmatrix} R(\theta_{t}) & \mathbf{x}_{t}  \\ 0_{1\times2} &  1 \end{bmatrix} \in SE(2) $, control input variable $ \mathcal{U}_{t} = \begin{bmatrix} [ \omega_{t} ] _ {\times} & \nu_{t}  \\ 0_{1\times2} &  1 \end{bmatrix} $, and process noise variable $ \mathcal{W}_{t} = \begin{bmatrix} [w_{t}^{\theta}]_{\times} & \mathbf{w}_{t}  \\ 0_{1\times2} &  0 \end{bmatrix} $, from \eqref{car_eq_2}, \eqref{car_eq_2}, and \eqref{rotation_eq}, the continuous state equation $ d\mathcal{X}_{t}/dt = \mathcal{X}_{t} \mathcal{U}_{t} + \mathcal{X}_{t} \mathcal{W}_{t} = f(\mathcal{X}_{t}, \mathcal{U}_{t}) + \mathcal{X}_{t} \mathcal{W}_{t} $ on the group can be obtained. According to Remark \ref{discretization_remark} and Lemma \ref{BCH_formula}, $ \mathcal{X}_{k+1} = \mathcal{X}_{k} \exp (\delta \mathcal{T}_{\mathcal{X}_{k}}(\mathcal{X}_{k} \mathcal{U}_{k} + \mathcal{X}_{k} \mathcal{W}_{k})) =  \mathcal{X}_{k} \exp(\delta \mathcal{U}_{k} + \delta \mathcal{W}_{k}) \approx \mathcal{X}_{k} \exp(\delta \mathcal{U}_{k}) \exp( \delta \mathcal{W}_{k}) = f(\mathcal{X}_{k}, \mathcal{U}_{k}) \exp(\delta \mathcal{W}_{k}) $ can be obtained, by ignoring the higher order infinitesimal terms $ O(\left \| \mathcal{U}_{k} \right \| ^ {2}, \left \| \mathcal{W}_{k} \right \| ^ {2}, \left \| \mathcal{U}_{k} \right \| \left \| \mathcal{W}_{k} \right \| ) $. 

$ \forall \mathcal{X}_{1}, \mathcal{X}_{2} \in SE(2) $, $ f(\mathcal{X}_{1}, \mathcal{U}_{k}) f(I_{d}, \mathcal{U}_{k})^{-1} f(\mathcal{X}_{2}, \mathcal{U}_k) = \mathcal{X}_{1} \exp(\delta \mathcal{U}_{k}) \exp( - \delta \mathcal{U}_{k}) \mathcal{X}_{2} \exp(\delta \mathcal{U}_{k}) = \mathcal{X}_{1} \mathcal{X}_{2} \exp(\delta \mathcal{U}_{k}) = f(\mathcal{X}_{1} \mathcal{X}_{2}, \mathcal{U}_k) $. The state equation $f(\mathcal{X}_{k}, \mathcal{U}_{k})$ in \eqref{Lie_matrix} satisfies the group affine property \eqref{group_affine_property}. Thus equation \eqref{Lie_matrix} is the system described in \eqref{sys_eq(a)}. 

\subsection*{C. Proof of the linear error dynamic}
Using Lemma \ref{Fundamental_property_of_invariant_filtering} and \eqref{Lie_matrix}, we have $ f(\hat{\mathcal{X}}_{k}, \mathcal{U}_{k})^{-1} f(\mathcal{X}_{k}, \mathcal{U}_{k}) \\= \exp(- \delta \mathcal{U}_{k}) \hat{\mathcal{X}}_{k}^{-1} \mathcal{X}_{k} \exp(\delta \mathcal{U}_{k}) = \exp(-  [\delta u_{k}]^{\land} ) \exp( [\epsilon_{k}^{L}]^{\land} ) \\ \exp( [\delta u_{k}]^{\land}) = (I_{3} - [\delta u_{k}]^{\land} + O(\cdotp) ) (I_{3} + [\epsilon_{k}^{L}]^{\land} + O(\cdotp)) (I_{3} + [\delta u_{k}]^{\land} + O(\cdotp)) = I_{3} + [ (I_{3} - \delta ad_{u_{k}}) \epsilon_{k}^{L} ]^{\land} + O(\cdotp) = \exp( [ (I_{3} - \delta ad_{u_{k}}) \epsilon_{k}^{L} ]^{\land} ) $, thus $ A_{k}^{L} = I_{3} - \delta ad_{u_{k}} $. Since the control input vector is $ u_{k} = \begin{bmatrix} \omega_{k} \\ \nu_{k}^{l} \\ 0 \end{bmatrix} $, then $ ad_{u_{k}} = \begin{bmatrix} 0 & 0 & 0 \\ 0 & 0 & -\omega_{k} \\ -\nu_{k}^{l} & \omega_{k} & 0 \end{bmatrix} $. For the experiments in this paper, it is assumed that the traveling speed and angular velocity of the vehicle remain constant, so $ A_{k}^{L} $ is a constant matrix. 

In order to make the output matrix $ C_{k}^{L} $ defined in Theorem \ref{Update_Law} and Lemma \ref{Measurement_linear_property} a constant matrix, an alternative innovation $ z_{k} = \hat{\mathcal{X}}_{k}^{-1} \mathcal{Y}_{k} - d $ is used. Thus, according to \eqref{car_meas_lie} and Remark \ref{Measurement_equation_modification}, we have
\begin{equation*}
	\begin{aligned}
		z_{k} &= \hat{\mathcal{X}}_{k}^{-1} \mathcal{X}_{k} d + \hat{\mathcal{X}}_{k}^{-1} d - d \\
		&= \exp([\epsilon_{k}^{L}]^{\land}) d - d + \hat{\mathcal{X}}_{k}^{-1} d \\
		&\approx [\epsilon_{k}^{L}]^{\land} d + \begin{bmatrix} R(\hat{\theta}_{k})^{\top} v_{k}  \\ 0 \end{bmatrix} \\
		&= \begin{bmatrix} 0 & 1 & 0 \\ 0 & 0 & 1 \\ 0 & 0 & 0 \end{bmatrix} \epsilon_{k}^{L} + \begin{bmatrix} R(\hat{\theta}_{k})^{\top} v_{k}  \\ 0 \end{bmatrix} \\
		&= C^{L} \epsilon_{k}^{L} + \begin{bmatrix} R(\hat{\theta}_{k})^{\top} v_{k}  \\ 0 \end{bmatrix} \\
		&= \tilde{p}^{\top} \tilde{C}^{L} \epsilon_{k}^{L} + \tilde{p}^{\top} R(\hat{\theta}_{k})^{\top} v_{k}
	\end{aligned}
\end{equation*}
where $ d = \begin{bmatrix} 0_{2\times1}  \\ 1 \end{bmatrix}$, $ \tilde{p} $, and $ \tilde{C}^{L} $ are defined as \eqref{InZSMF_Error}. And $ \epsilon_{k}^{L} = [ \sigma \; u_{1} \; u_{2}] ^{\top} \in \mathbb{R}^{3} $ is defined as Appendix A. 

It can be known from the proof of Theorem \ref{Update_Law} that $ \exp([\epsilon_{k+1}^{L}]^{\land}) = \eta_{k+1}^{L} = \exp([A_{k}^{L}\epsilon_{k}^{L} + \delta w_{k} - L_{k}z_{k}]^{\land}) = \exp([A_{k}^{L}\epsilon_{k}^{L} + \delta w_{k} - L_{k}\tilde{p}^{\top} \tilde{C}^{L} \epsilon_{k}^{L} - L_{k} \tilde{p}^{\top} R(\hat{\theta}_{k})^{\top} v_{k}]^{\land}) $. Let $ L_{k} = \tilde{L}_{k} \tilde{p} $, then $ \epsilon_{k+1}^{L} = (A_{k}^{L} - \tilde{L}_{k} \tilde{C}^{L})\epsilon_{k}^{L} + \delta w_{k} - \tilde{L}_{k} R(\hat{\theta}_{k})^{\top} v_{k} $ can be obtained. 

\section*{References}
\bibliographystyle{IEEEtran}
\bibliography{IEEE_ref}

\end{document}